\documentclass[12pt]{iopart}
\usepackage{iopams}  
\usepackage{graphicx}
\usepackage{dsfont}
\newcommand{\vac}{\varnothing}
\newcommand{\ket}[1]{|#1\rangle}
\newcommand{\bra}[1]{\langle#1|}

\begin{document}

\title[]{Non-hermitian Hamiltonian description for quantum plasmonics: from dissipative dressed atom picture to Fano states} 

\author{H. Varguet$^1$, B. Rousseaux$^{2,3}$, D. Dzsotjan$^{4}$, H. R. Jauslin$^1$, S. Gu\'erin, G. Colas des Francs$^1$}

\address{$^1$Laboratoire Interdisciplinaire Carnot de Bourgogne, CNRS UMR 6303, Universit\'e Bourgogne Franche Comt\'e,
	BP 47870, 21078 Dijon, France}
\address{$^2$Department of Physics, Chalmers University of Technology, 412 96 G\"oteborg, Sweden.}
\address{$^3$Department of Microtechnology and Nanoscience – MC2, Chalmers University of Technology, 412 96 G\"oteborg, Sweden.}
\address{$^4$ Wigner Research Center for Physics, Hungarian Academy of Sciences, Konkoly-Thege Miklos ut 29-33, H-1121 Budapest, Hungary}
\ead{gerard.colas-des-francs@u-bourgogne.fr}

\begin{abstract}
We derive effective Hamiltonians for a single dipolar emitter coupled to a metal nanoparticle (MNP) with particular attention devoted to the role of losses. For small particles sizes, absorption dominates and a non hermitian effective Hamiltonian describes the dynamics of the hybrid emitter-MNP nanosource. We discuss the coupled system dynamics in the weak and strong coupling regimes offering a simple understanding of the energy exchange, including radiative and non radiative processes. We define the plasmon Purcell factors for each mode. For large particle sizes, radiative leakages can significantly perturbate the coupling process. We propose an effective Fano Hamiltonian including plasmon leakages and discuss the link with the quasi-normal mode description. We also propose Lindblad equations for each situation and introduce a collective dissipator for describing the Fano behaviour.
\end{abstract}

%
%
%
%
%

\section{Introduction} 
Cavity quantum electrodynamics (cQED) takes benefit from the long duration of the light-matter interaction in optical microcavities. This has opened the door to important applications including low threshold laser \cite{Nomura-Arakawa:2008}, supercontinuum laser \cite{Grelu:2016} or indistinguishable single photon sources \cite{Laurent-Abram:2005}. In the strong coupling regime, the Jaynes-Cummings ladder anharmonicity can lead to photon blockade \cite{Faraon-Vukovic:08} and  the coherence of the hybrid polariton states permits the realization of low power laser \cite{Savvidis:14}. Optical microcavities present extremely high quality factors but at the price of diffraction limited sizes, limiting integration capabilities. It is therefore of strong interest to transpose cQED to nanophotonics and plasmonics \cite{Paspalakis-Vitanov:09,Cuche2010,Agio:2012,Tame-Maier:2013,Benson:2014,GCF-Barthes-Girard:2016,Marquier-Sauvan-Greffet:2017,Vasa-Lienau:17}. Particular attention has been devoted to the strong coupling regime \cite{Truegler-Hohenester:2008,AbreeGuebrou-Belessa:2012,Delga-GarciaVidal:2014,Zengi-Kall-Shegai:2015,Kewes-Benson:2018} since it offers the possibility of the control of the dynamics of the light emission, as {\it e.g.} photon blockade \cite{Smolyaninov-Zayats-Gungor-Davis:2002,Alpeggiani-Gerace:2016} or coherent control \cite{Fleischhauer:2010,Rousseaux-GCF:2016,SaezBlaquez-GarciaVidal:17}. Moreover, quadrupolar forbidden atomic transitions can occur in plasmonic cavities thanks to the strong plasmon field gradient \cite{Kern-Martin:12,CuarteroGonzalez:2018}. In addition, in the weak coupling regime, the acceleration of single photon source cadency by coupling to plasmons opens the doors to high operation speed quantum functionnalities beyond the limited bandwidth (high Q) of optical microcavities systems \cite{Tame-Maier:2013,Schietinger-Barth-Aichele-Benson:2009,Marty-Arbouet-Paillard-Girard-GCF:2010,Singh-vanHulst:18}.

Quantum plasmonic systems behave like open quantum systems because of strong losses originating from absorption into the metal or from radiation leakages to the far-field. The dynamics of open quantum systems can be described considering either a master equation or a non-hermitian effective Hamiltonian. Generally, the master equation is derived from an hermitian Hamiltonian by tracing out the baths into which the energy is lost. An non-hermitian effective Hamiltonian can also be derived describing the full dynamics of the same open system except the ground state of the system. More precisely, the non hermitian Hamiltonian lacks the feeding term ({\it e.g.} laser pump) appearing in the master equation but is fully equivalent when no pumping occurs \cite{Visser-Nienhuis:95}. Compared to the master equation approach, the use of a non hermitian Hamiltonians strongly reduces the required numerical ressources when the number of atomic and plasmonic states increase. In addition, it is worthwhile to note that it is difficult to separate the plasmon contribution from the absorption and radiation baths so that deriving a master equation can be delicate. However, a master equation can be inferred from the non hermitian effective Hamiltonian on the basis of its similarity with other open quantum systems. Therefore, we investigate in detail the properties of the non-hermitian effective Hamiltonian we recently derived for localized surface plasmons (LSPs) \cite{Rousseaux-GCF:2016,Dzsotjan-GCF:2016}. We specifically discuss the role of losses in the effective Hamiltonian construction, leading to non-hermitian behaviours. LSPs constitute a benchmarch for investigating non-hermitian behaviours by an analytical description and by experimental direct characterization of their response. For instance, self-hybridization of LSPs of different orders due to non-hermiticity has been recently demonstrated \cite{Kociak:18}. In the following, we investigate the dynamics of hybrid nanosources on the basis of the eigenmodes of the non-hermitian Hamiltonian and discuss the link with quasi-normal modes (QNM) approaches \cite{Sauvan-Lalanne:2013,ZambranaPuyalto-Bonod:2015,LalanneReview:18}.

In section \ref{sect:LSP-QED}, we recall the main steps leading to the definition of an effective Hamiltonian. We identify the coupling strength and illustrate the procedure considering a quantum emitter coupled to a silver nanoparticle. We demonstrate that the  hybrid nanosource optical response can be described in full analogy with a cQED representation where dissipative localized surface plasmons (LSP) play a role analogous to leaky cavity modes. The strong and weak coupling regimes are discussed in section \ref{sect:DressedAtom} and \ref{sect:weak}, respectively. 
We finally investigate in section \ref{sect:Fano} the impact of LSP leakage on the effective Hamiltonian structure, notably by introducing Fano states that originates from coupling the LSP discrete states to the free-space continuum. For this purpose, we start from a bath model inferred from the effective Hamiltonian derived in \S \ref{sect:LSP-QED} and discuss the role of the leakages in this model. 

\section{LSP field quantization and effective model}
\label{sect:LSP-QED}
\begin{figure}[h!]
	\hspace{0.5cm}
\includegraphics[width=0.4\textwidth]{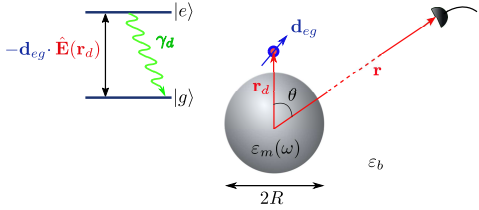}
	\vspace{0.5cm}
	\caption{Scheme of the hybrid system embedded in a background material with permittivity $\varepsilon_b$. A spherical MNP of radius $R$ and permittivity $\varepsilon_m(\omega)$ is coupled to a TLS dipolar emitter.  $\mathbf{r}_d$ and  $\mathbf{r}$ refer to the emitter and detector positions respectively. The inset describes the TLS system.}
\label{MNP}
\end{figure}

We represent in Fig. \ref{MNP} the hybrid system that consists of a dipolar emitter close to a metal nanoparticle (MNP). The dielectric constant of the background medium is $\varepsilon_b$. The dipolar quantum emitter is a two-level system (TLS) with ground and excited states $\ket{g}$ and $\ket{e}$ of energy $\hbar \omega_g$ and $\hbar \omega_e$, respectively. The dipole moment of the optical transition is denoted by $\mathbf{d}_{eg}$. The decay rate of the excited state is denoted by $\gamma_0=\gamma_0^{rad}+\gamma_0^{NR}$, including the radiative and intrinsic non radiative contributions. $\gamma_0^{rad}$ is the radiative contribution in the homogeneous medium of optical index $n_b=\sqrt{\varepsilon_b}$
\begin{equation}
\gamma_0^{rad}=n_b\frac{d_{eg}^2\omega_0^3}{3\pi\epsilon_0 \hbar c^3} 
\label{eq:Grad0}
\end{equation}
and we define the intrinsic quantum yield $\eta=\gamma_0^{rad}/\gamma_0$.
The MNP is characterized by the dielectric constant $\varepsilon_m(\omega)=\varepsilon_R(\omega)+i\varepsilon_I(\omega)$ that can be extracted from tabulated data or modelled with a Drude model. Without loss of generality, we assume a Drude-like
behavior $\varepsilon_m(\omega)=\varepsilon_{\infty}-\omega_p^2/(\omega^2+i\Gamma_p\omega)$ with $\varepsilon_{\infty}=6$, $\hbar\omega_p=7.90$ eV and $\hbar\Gamma_p=51$ meV for silver. 
%

The Hamiltonian of the coupled system reads
\begin{eqnarray}
\hat H=&\hbar\omega_0\hat{\sigma}_{ee}-i\hbar\frac{\gamma_0}{2}\hat{\sigma}_{ee}
+\int d\mathbf{r} \int_0^{+\infty}\!\!\!\!\!\!\! d\omega\ \hbar\omega\  
\hat{\mathbf{f}}_\omega^{\dagger}(\mathbf{r})\cdot\hat{\mathbf{f}}_\omega(\mathbf{r})\nonumber\\
&-\left[\hat{\sigma}_{eg} \int_0^{+\infty}\!\!\!\!\!\!\! d\omega\ \mathbf{d}_{eg}\cdot\hat{\mathbf{E}}^+_\omega(\mathbf{r}_d)+H.c.\right]\label{hamil}.
\end{eqnarray}
$\omega_0=\omega_e-\omega_g$ is the transition angular frequency and we introduce the excited state population and raising operators of the emitter $\hat{\sigma}_{ee}=\ket{e}\bra{e}$ and 
$\hat{\sigma}_{eg}=\ket{e}\bra{g}$, respectively. In equation (\ref{hamil}), the first term refers to the TLS energy and we have phenomelogically introduced the decay rate $\gamma_0$ of the excited state. 
The third term describes the total energy of the electromagnetic field where $\hat{\mathbf{f}}^{\dagger}({\mathbf r})$ [$\hat{\mathbf{f}}({\mathbf r})$] is the  polaritonic vector field operator at the position ${\mathbf r}$ associated to the creation (annihilation) of a quantum in the presence of the MNP.
The last term describes the emitter-field interaction under the rotating-wave approximation. 

The electromagnetic field must be quantized by taking into account the dispersing and absorbing nature of the metal \cite{Knoll-Schell-Welsch:2001,DrezetPRA:16,DrezetPRA:17}. Within the Langevin type model of ref. \cite{Knoll-Schell-Welsch:2001}, the electric field operator can be written as
\begin{eqnarray}
\mathbf{\hat{E}}^+_\omega(\mathbf{r})=& i\sqrt{\frac{\hbar}{\pi\epsilon_0}}k_0^2
\int d{\mathbf{r}'} \sqrt{\varepsilon''_\omega(\mathbf{r}')}
{\mathbf G}_\omega({\mathbf r},{\mathbf r}')\hat{{\mathbf f}}_\omega({\mathbf r'}),
\label{eq:OpE}
\end{eqnarray}
where $k_0=\omega/c$ is the wavenumber and $\mathbf{G}({\mathbf r},{\mathbf r}')$ is the Green tensor associated to the electric field response at position ${\mathbf r}$ from an excitation localized at ${\mathbf r}'$ in the medium. This expressions fails in describing the electric field operator in free-space (for which $\varepsilon''=0$). Recent works discuss a general definition of the electric field operator including the free-space contribution \cite{Dorier-Jauslin:18}. 

In the following, we investigate the optical response of the emitter-MNP coupled system. 
The wave function of the hybrid system can be written at time $t$ as
\begin{eqnarray}
&&\ket{\psi(t)} = \ C_e (t)e^{-i\omega_0 t}\ket{e,\varnothing}
+ \int d\mathbf{r}\int_0^{\infty}d \omega\ e^{-i\omega t}\mathbf{C}_{\omega}(\mathbf{r},t)\cdot\ket{g,\mathbf{1}_\omega(\mathbf{r})}  \;,
\label{wavefun}
\end{eqnarray}
where $\ket{e,\varnothing}$ refers to the emitter in its excited state and no LSP mode excited whereas
$\ket{g,\mathbf{1}_\omega(\mathbf{r})}$ refers to the emitter in its ground state and a single excited polariton of energy $\hbar \omega$. The elementary excitation at position $\mathbf{r}$ is defined through the action of the bosonic vector field operator on the vacuum state $\mathbf{f}_\omega^{\dag}(\mathbf{r})\ket{\varnothing}=\ket{\mathbf{1}_\omega(\mathbf{r})}$.

Up to know, we consider a continuous description for polaritons. However, the dynamics of the coupled system deserve attention regarding the excitation of LSP modes of the MNP. The Green tensor ${\mathbf G}$ governs this dynamics and contains all the modal informations of the MNP in terms of the Mie expansion

\begin{eqnarray}
\label{eq:En}
&&\mathbf{\hat{E}}^+_\omega(\mathbf{r})=\sum_{n=1}^{\infty}\mathbf{\hat{E}}^+_{\omega,n}(\mathbf{r}) \;,  \\
\nonumber
&&\mathbf{\hat{E}}^+_{\omega,n}(\mathbf{r})= i\sqrt{\frac{\hbar}{\pi\epsilon_0}}k_0^2
\int d{\mathbf{r}'} \sqrt{\varepsilon''_\omega(\mathbf{r}')} {\mathbf G}_{\omega,n}({\mathbf r},{\mathbf r}')\hat{{\mathbf f}}_\omega({\mathbf r'}) \; \\
\nonumber
&&{\mathbf G}_\omega({\mathbf r},{\mathbf r}')=\sum_{n=1}^{\infty}{\mathbf G}_{\omega,n}({\mathbf r},{\mathbf r}') \;,
\end{eqnarray}
where ${\mathbf G}_n$ refers to the contribution of the $n^{th}$ plasmon LSP$_n$ (dipolar plasmon for $n=1$, quadrupolar plasmon for $n=2$, {\it etc.}) so that $\mathbf{\hat{E}}_{\omega,n}$ is the electric field operator associated to the LSP$_n$ mode.  This leads us to define the bosonic creation operator for a given position of the emitter, satisfying the commutation relation $[\hat{a}_{\omega',n'}(\mathbf{r}_d),\hat{a}_{\omega,n}^{\dagger}(\mathbf{r}_d)]=\delta(\omega-\omega')\delta_{n,n'}$ \cite{Rousseaux-GCF:2016,Dzsotjan-GCF:2016} 
\begin{eqnarray}
&&\hat{a}_{\omega,n}(\mathbf{r}_d)=\frac{1}{i\hbar\kappa_{\omega,n}(\mathbf{r}_d)} \mathbf{d}_{eg}\cdot \mathbf{\hat E}^+_{\omega,n}(\mathbf{r}_d) \;, \\
&&|\kappa_{\omega,n}(\mathbf{r}_d)|^2=\frac{k_0^2}{\hbar\pi\epsilon_0} Im\left[ \mathbf{d}_{eg}\cdot \mathbf{G}_{\omega,n}(\mathbf{r}_d,\mathbf{r}_d) \cdot \mathbf{d}_{eg}^\star \right ] \;\label{lien}
\end{eqnarray}

The excitation at the frequency $\omega$ of a single plasmon of order $n$ is $\ket{1_{\omega,n}(\mathbf{r}_d)}=\hat{a}_{\omega,n}^{\dagger}(\mathbf{r}_d)\ket{\varnothing}$. Moreover $\kappa_{\omega,n}$ is the emitter-LSP$_n$ coupling which is the key parameter to build the effective model. Truncating the modal decomposition to the number $N$ of modes involved in the coupling process, the full Hamiltonian (Eq. \ref{hamil}) becomes (see ref. \cite{Dzsotjan-GCF:2016,Castellini:18} for details)
\begin{eqnarray}
\hat H=\hbar\omega_0\hat{\sigma}_{ee}-i\hbar\frac{\gamma_0}{2}\hat{\sigma}_{ee}
+\sum_{n=1}^N\int_0^{+\infty}\!\!\!\!\!\!\! d\omega \hbar \omega \hat{a}^{\dagger}_{\omega,n}(\mathbf{r}_d)\hat{a}_{\omega,n}(\mathbf{r}_d) \\
+i\hbar\sum_{n=1}^N\int_0^{+\infty}\!\!\!\!\!\!\! d\omega\kappa_{\omega,n}^*(\mathbf{r}_d)\hat{a}^{\dagger}_{\omega,n}(\mathbf{r}_d)\hat{\sigma}_{ge}-H.c. \nonumber
\end{eqnarray}

\begin{figure}
\includegraphics[width=12cm]{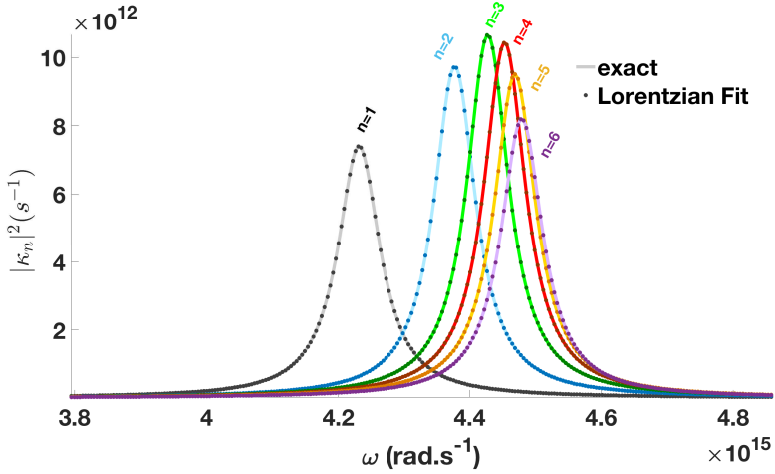}
\caption{Coupling constant spectra for the six first LSP$_n$ modes calculated using Eq. (\ref{lien}). The emitter is located 2nm from the silver MNP. The lorentzian fits follow eq. (\ref{structuration}).}
\label{fig:LorentzianMNP}
\end{figure}
To finalize the effective model, we take benefit from the  Lorentzian profile for each resonance. In Fig. \ref{fig:LorentzianMNP}, we plot the coupling constant $\vert \kappa_{\omega,n}\vert^2$ for several modes. We observe an excellent agreement with a Lorentzian profile so that $\kappa_{\omega,n}$ can be written as
\begin{eqnarray}
\kappa_{\omega,n}(\mathbf{r}_d)=\sqrt{\frac{\Gamma_n}{2\pi}}\frac{i g_n(\mathbf{r}_d)}{\omega-\omega_n+i\frac{\Gamma_n}{2}} \;. \label{structuration}
\end{eqnarray}
$\omega_n$ and $\Gamma_n$ are the mode resonance frequency and width, respectively. These parameters are deduced from a Lorentzian fit, but analytical expresssions are also available in the  near-field regime, revealing the radiative and non radiative contributions to the mode's rate of losses $\Gamma_n$ (see \ref{sect:AnnexRate}).

\begin{figure}
\includegraphics[width=14cm]{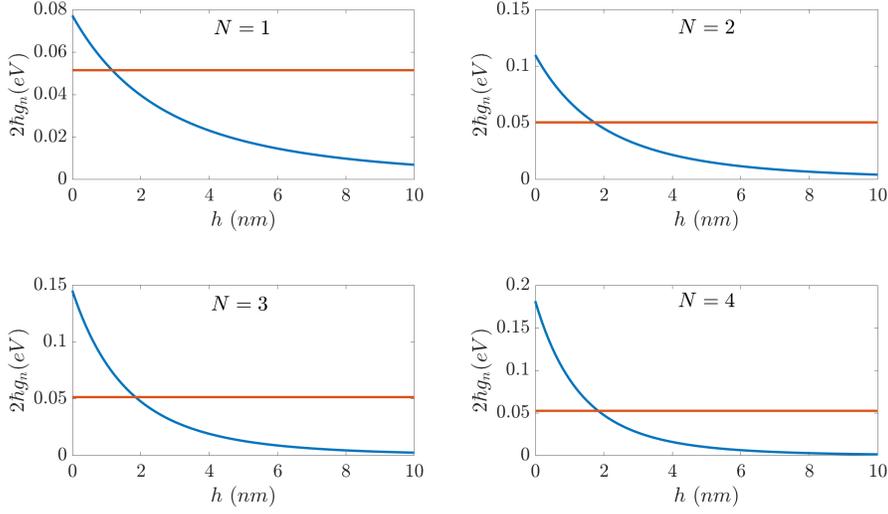}
\caption{Coupling strength $2\hbar g_n$ to the first LSP$_n$ modes (n=1,\ldots, 4) as a function of the distance to the particle surface.  The horizontal line represents the  LSP$_n$ losses $\hbar\Gamma_n$.}
\label{fig:Gn}
\end{figure}

Finally, the effective Hamiltonian is obtained by integrating over the angular frequency $\omega$ in order to establish a set of $N$ discrete modes. To this purpose, we define the $n^{th}$ plasmonic operator 
\begin{eqnarray}
&&\hat{a}_{n}(\mathbf{r}_d) =\frac{1}{i g_n(\mathbf{r}_d)}\int_0^{+\infty}\!\!\!\!\!\!\! d\omega \kappa_{\omega,n}(\mathbf{r}_d)\hat{a}_{\omega,n}(\mathbf{r}_d)
\end{eqnarray}
and the wavefunction of the hybrid system takes the form
\begin{eqnarray}
&&\ket{\psi_{eff}(t)} =  C_e (t)\ket{e,\varnothing} 
+\sum_{n=1}^N\mathbf{C}_n(t)\cdot\ket{g,\mathbf{1}_n(\mathbf{r}_d)} \;, 
\label{eq:PsiEff}
\end{eqnarray} 
such that $\ket{1_{n}(\mathbf{r}_d)}=\hat{a}_{n}^{\dagger}(\mathbf{r}_d)\ket{\varnothing}$ defines the excitation of a single plasmon LSP$_n$. This effective wavefunction lets explicitely appear the contribution of the TLS excited state and all plasmon modes.
We explicitely indicate the dependence on the emitter position $\mathbf{r}_d$ to indicate that the parameters of the effective model are defined for a given position of the atomic system. Since we are considering a single emitter coupled to the MNP, we can safely omit the explicit dependence on $\mathbf{r}_d$ in the following. Finally, we define the effective Hamiltonian $H_{eff}$ so that $i\hbar \partial_t \ket{\Psi_{eff}(t)}=\hat H_{eff} \ket{\Psi_{eff}(t)}$. Identifying the dynamics of the wavefunctions in the discrete and continuum descriptions, we obtain the matrix representation of the effective Hamiltonian in the basis $\{\vert e,\vac\rangle,\vert g, 1_1\rangle,\cdots,\vert g, 1_N\rangle \}$  \cite{Dzsotjan-GCF:2016,Castellini:18}
\begin{eqnarray}
H_{eff}=\hbar \left(
\begin{array}{ccccc}
-i\frac{\gamma_0}{2} & g_1 & g_2 & \cdots & g_N\\
g_1 & \Delta_1-i\frac{\Gamma_1}{2} & 0 & \cdots & 0\\
g_2 & 0 & \Delta_2-i\frac{\Gamma_2}{2} & \ddots & \vdots\\
\vdots & \vdots & \ddots & \ddots & 0\\
g_N & 0 & \cdots & 0 & \Delta_N-i\frac{\Gamma_N}{2}
\end{array}
\right)
\,,
\label{eff_hamil} 
\end{eqnarray}
where $\Delta_n=\omega_n-\omega_{0}$ is the detuning of the LSP$_n$ resonance from the TLS emission. The effective Hamiltonian describe the evolution of a sub-system (atomic states + plasmons) of a larger configuration (atom+phonon bath+radiation bath) so that it is non-hermitian and presents losses on its diagonal.
This effective Hamiltonian provides a very practical representation of the hybrid configuration, presenting a direct analogy with a cQED system. $g_n$ defines the coupling strength of the emitter to the MNP $n^{th}$ mode. $\omega_n$ and $\Gamma_n$ are the LSP$_n$ frequency and rate of losses, respectively. $\omega_n,\Gamma_n$ and $g_n$ depend on the MNP material and its size but the coupling strength $g_n$ depends also on the distance to the MNP. These parameters are deduced from a Lorentzian fit of the coupling constant $\kappa_{\omega,n}$, as presented in Fig. \ref{fig:LorentzianMNP}. We plot on Fig. \ref{fig:Gn} the coupling strength to the LSP$_n$ (n=1 to 4) modes as a function of the distance. We superimpose the mode losses rate $\Gamma_n$. We observe that mode losses are governed by Joule losses in the metal ($\hbar \Gamma_n \approx \hbar\Gamma_p = 51$ meV).  For larger particles, the dipolar mode losses are larger due to higher radiative losses. This will lead us to propose a modified Fano effective Hamiltonian in section \ref{sect:Fano}.

Finally, the coupling strength ($\hbar g_n$) can overcome the plasmon losses ($\hbar \Gamma_n \approx 51$ meV) and TLS losses ($\hbar \gamma_0 = 15$ meV)  at very short distances so that a strong coupling regime occurs \cite{Varguet-GCF:16}. Note that higher coupling strength can be expected for smaller separation distances. However, we have to keep in mind that for distances below 1 nm, non local effects can occur and the dielectric function presents a $k$ dependence $\epsilon_m(k,\omega)$, not taken into account here, that can screen the coupling strength \cite{Leung:1990,Castanie-Boffety-Carminati:2010,Girard-Cuche:2015}.

\begin{figure}
\includegraphics[width=8cm]{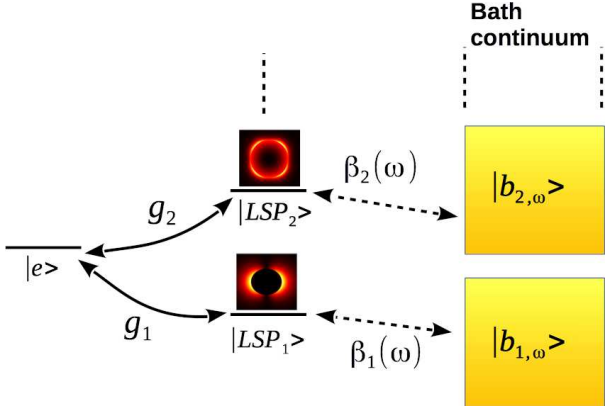}
\caption{Open quantum system presenting a dynamics equivalent to one of the effective non hermitian Hamiltonian Eq. (\ref{eff_hamil}).}
\label{fig:BathcQED}
\end{figure}

It is worth noticing that this effective Hamiltonian presents a one to one mapping with a non hermitian Hamiltonian of cQED. By analogy to the cQED treatment, we can describe LSPs dissipation by the coupling to a continuum bath (see figure \ref{fig:BathcQED}). We separate the system $\mathcal{S}$ composed of emitter and LSPs from the environment $\mathcal{E}$ associated to the bath. Hence, we introduce a new Hamiltonian leading to the same dynamics as the original one
\begin{eqnarray}
\label{H_SE}
\hat{H}_{SE}&=&H_S+H_E+H_I,\\
\nonumber
H_S&=&\sum_{n=1}^{N}\hbar\Delta_n\hat{a}_n^{\dagger}\hat{a}_n+\hbar\sum_{n=1}^{N}\left(g_n\hat{\sigma}_{eg}\hat{a}_n+g_n\hat{a}_n^{\dagger}\hat{\sigma}_{ge}\right),
\\
\nonumber
H_E&=&\int d\omega\,\hbar\omega\sum_{n=1}^{N}\hat{b}_{\omega,n}^\dagger\hat{b}_{\omega,n},\\
\nonumber
H_{I}&=&i\hbar\int d\omega\,\sum_{n=1}^{N}\beta_n(\omega)\left(\hat{b}_{\omega,n}^{\dagger}\hat{a}_n-\hat{a}_{n}^{\dagger}\hat{b}_{\omega,n}\right).
\end{eqnarray}
The system Hamiltonian $H_S$ describes the interaction between the emitters and the LSPs and is hermitian. The environment Hamiltonian $H_E$ involves all the bath oscillators of energies $\hbar\omega$. For each cavity pseudo-mode of order $n$ we define a reservoir. 
The interaction between the system and the environment is described by the Hamiltonian $H_I$ where $\zeta_n(\omega)$ characterizes the coupling between the pseudo-modes and their associated reservoirs. For flat coupling $\beta_n(\omega)\approx \beta_n(\omega_n)=\sqrt{\Gamma_n/2\pi}$,  this new Hamiltonian leads to the following Lindblad equation \cite{Hummer-Garciavidal:2013,Rousseaux-Shegai:18,Hughes-Knorr:18,Varguet-GCF:18}
\begin{eqnarray}
\frac{d\hat{\rho}(t)}{dt}=\frac{1}{i\hbar} \left[\hat{H}_S,\hat{\rho}(t)\right]+\mathcal{D}_0\left[\hat{\rho}(t)\right] +\mathcal{D}_{LSP}\left[\hat{\rho}(t)\right] 
\label{Lindblad}
\end{eqnarray}
where $\hat{\rho}$ is the density matrix of the emitter-LSPs system. The LSP dissipator naturally appears and is of the form
\begin{eqnarray}
\label{Dissipator}
&&\mathcal{D}_{LSP}\left[\hat{\rho}(t)\right]=-\sum_{n=1}^N 
\frac{\Gamma_n}{2}\left[\hat{a}^{\dagger}_{n}\hat{a}_{n}\hat{\rho}(t)+\hat{\rho}(t)\,\hat{a}^{\dagger}_{n}\hat{a}_{n}-2\hat{a}_{n} \hat{\rho}(t)\,\hat{a}^{\dagger}_{n}\right] \;.
\end{eqnarray}
but we phenomenologically introduce the emitter dissipator 
\begin{eqnarray}
\label{Dissipator0}
&&\mathcal{D}_0\left[\hat{\rho}(t)\right]=
-\frac{\gamma_0}{2}\left[\hat{\sigma}_{eg}\hat{\sigma}_{ge}\hat{\rho}(t)+\hat{\rho}(t)\,\hat{\sigma}_{eg}\hat{\sigma}_{ge}-2\hat{\sigma}_{ge}\hat{\rho}(t)\hat{\sigma}_{eg}\right] \,.
\end{eqnarray}
The direct derivation of $\mathcal{D}_0$ will be discussed in \S \ref{sect:Fano}.

The master equation (\ref{Lindblad}) is a model for the same dynamics as the non hermitian effective Hamiltonian (\ref{eff_hamil}) but also include the dynamics of the fundamental state $\vert g,\vac\rangle$. However, it is worth noticing that for the $N$ LSPs + 1 emitter states, the Lindblad master equation operates in a space of dimension $(N+1)^2$ whereas the effective Hamiltonian work in a space of dimension $(N+1)$, so that the effective Hamiltonian should be priviledged when possible.

\section{Dissipative dressed atom picture}
\label{sect:DressedAtom}
Considering all the $N$ LSP modes plus the TLS excited state in the effective Hamiltonian (Eq. \ref{eff_hamil}), we define $N+1$ hybrid modes that are the eigenvectors of $H_{eff}$.We denote their complex angular frequency by
\begin{eqnarray}
\lambda_m=\omega_m-i\frac{\gamma_m}{2} \,, (m=1,\ldots , 26)
\end{eqnarray}
where $\lambda_m$ is the eigenvalue of the effective Hamiltonian.

For such dissipative systems, we have to define right and left eigenvectors $\vert\Pi^R_m\rangle$ and $\vert\Pi^L_m\rangle$, respectively, satisfying $H_{eff}\vert\Pi^R_m\rangle=\lambda_m \vert\Pi^R_m\rangle$ and $H_{eff}^\dagger\vert\Pi^L_m\rangle=\lambda_m^\star \vert\Pi^L_m\rangle$, $\langle\Pi^L_m\vert\Pi^R_m\rangle=\delta_{mn}$. For a Hamiltonian of the form (\ref{eff_hamil}), one can simply connect them as follows (see \ref{sect:Biorthog}) \cite{Guerin:2010}
\begin{eqnarray}
\vert\Pi_m^R\rangle&=m_0\vert e,\varnothing\rangle+\sum_{n=1}^N m_n \vert g, 1_n\rangle, \\
\vert\Pi_m^L\rangle&=-m_0^\star\vert e,\varnothing\rangle+\sum_{n=1}^N m_n^\star \vert g, 1_n\rangle,
\label{eq:LeftVect0}
\end{eqnarray}
where $m_0$ and $m_n$ gives the weight of each mode $\vert e, \varnothing \rangle$ or $\vert g, 1_n\rangle$. 

\begin{figure}[h!]
	\includegraphics[width=0.5\textwidth]{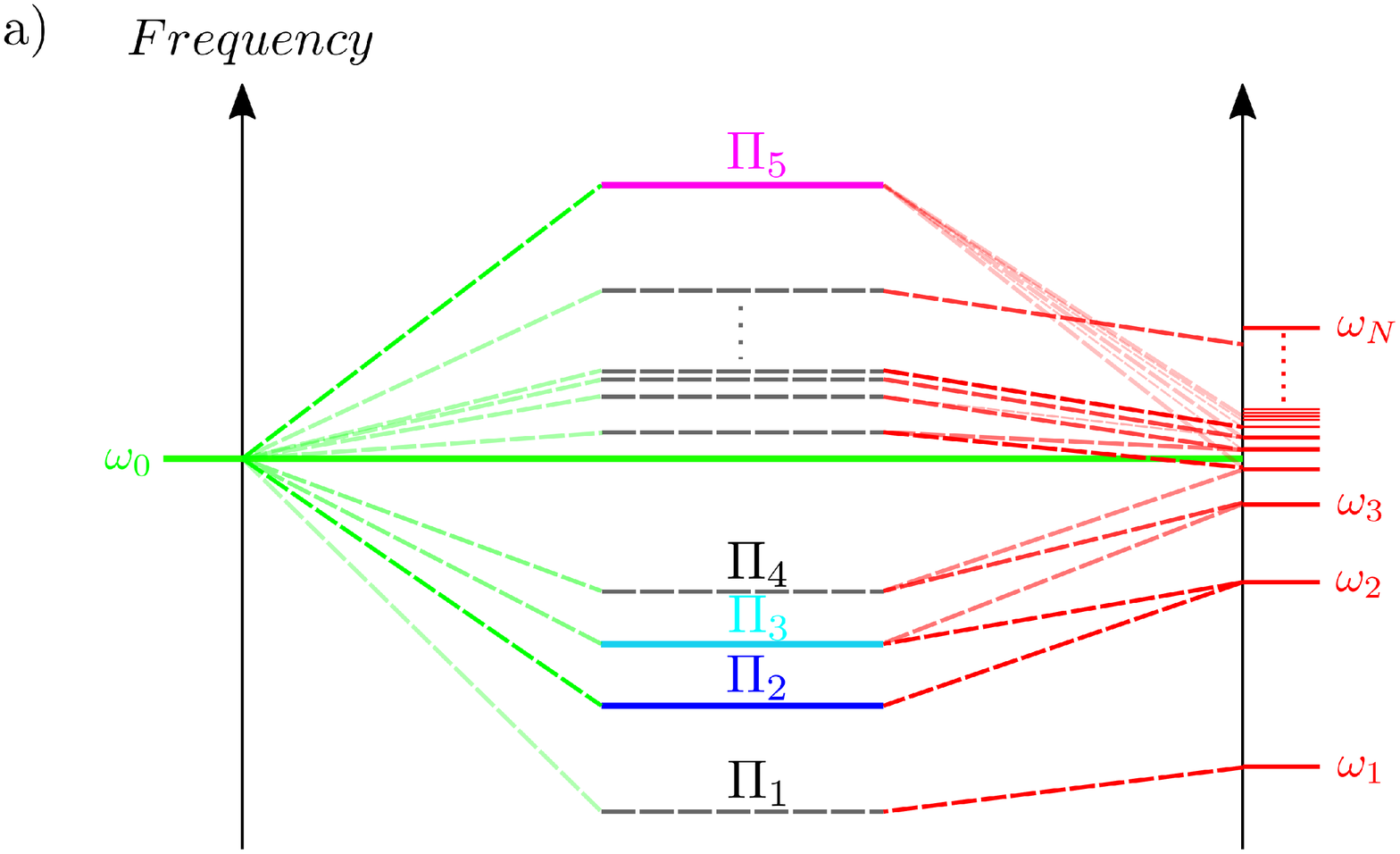}
	\includegraphics[width=0.42\textwidth]{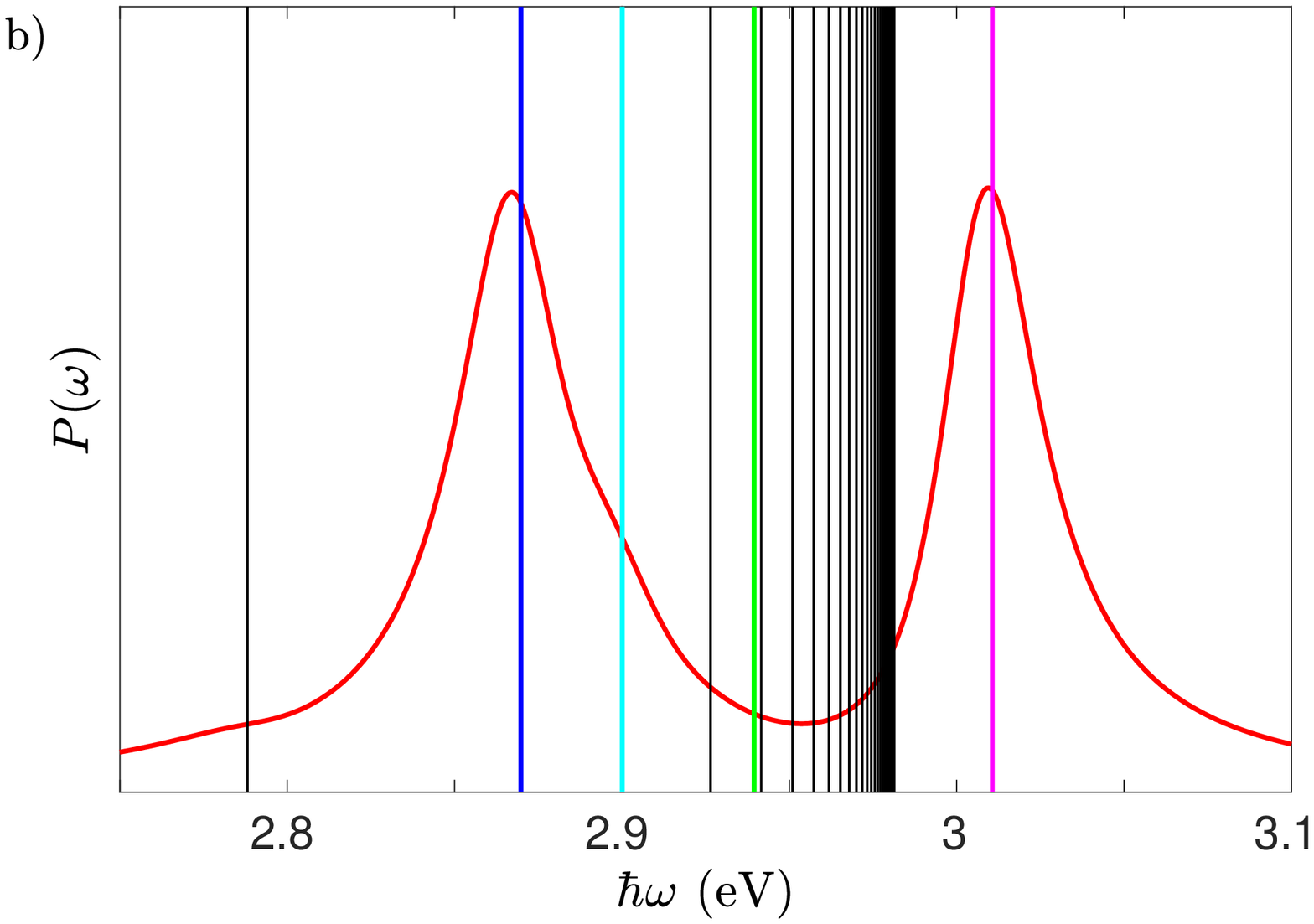}
	\caption{(a) Energy diagram of the dressed states. A thicker line indicates a stronger weight of the atom  ($\vert e,\varnothing\rangle$, left part) or LSP$_n$ mode ($\vert g, 1_n\rangle$, right part of the diagram). (b) Polarization spectrum. Black lines indicate the dressed states frequencies. In (b) the green line corresponds to the emission frequency of the emitter ($\hbar\omega_{0}=2.94$ eV) leading to the strong coupling. The blue (magenta) line refers to the frequency $\Omega_2$ ($\Omega_5$) of the dressed state $\Pi_2$ ($\Pi_5$). The cyan line near $\hbar\omega\approx 2.9$ eV indicates the frequency of the $\Pi_3$ state. Adapted from \cite{Varguet-GCF:16}. \label{degen_multi}}
\end{figure}

Finally, the wavefunction (expression \ref{eq:PsiEff}) can be represented at time $t$ by:  
\begin{eqnarray}
\ket{\psi_{eff}(t)}=C_e (t)\ket{e,\varnothing}
+\sum_{n=1}^N\mathbf{C}_n(t)\cdot\ket{g,\mathbf{1}_n}
= \sum_{m=1}^{N+1} \eta_m\vert \Pi_m^R \rangle e^{-i\lambda_m t} \,,
\label{eq:PsiDyn}
\end{eqnarray}
with  $\eta_m=\langle\Pi_m^L \vert \psi(0)\rangle=-m_0$ if we assume an emitter initially in its excited state and no LSP mode populated. Therefore the hybrid system wavefunction is expanded on the non hermitian Hamiltonian eigenmodes defining the atomic states dressed by LSP modes, as depicted in Fig. \ref{degen_multi}a)\cite{Varguet-GCF:16} . It defines a Jaynes-Cummings ladder, that has a same form as a cQED model, and that can be probed considering the near-field emission spectrum. Indeed, for an emitter initially in its excited state, the polarization spectrum takes the form 
\begin{eqnarray}
\nonumber
P(\omega)&=&\left| \int_0^\infty dt e^{i(\omega-\omega_0) t}C_e(t) \right|^2 \,, \\
&=&\left| \sum_{m=1}^{N+1} \frac{m_0^2}{\omega -(\omega_0+\omega_m) +i\frac{\gamma_m}{2}}\right|^2 \,.
\label{eq:SpectraHeff}
\end{eqnarray} 
If the dressed states are well separated in energy, the near-field spectrum is approximated by a sum of Lorentzians peaked at their resonance energy and with a FWHM $\gamma_m$. In the present case, some of the dressed states are not sufficiently separated so that the exact expression (eq. \ref{eq:SpectraHeff}) has to be used but still the effective model gives a clear understanding of the Rabi splitting (see Fig. \ref{degen_multi}b).

\begin{figure}
\includegraphics[width=8cm]{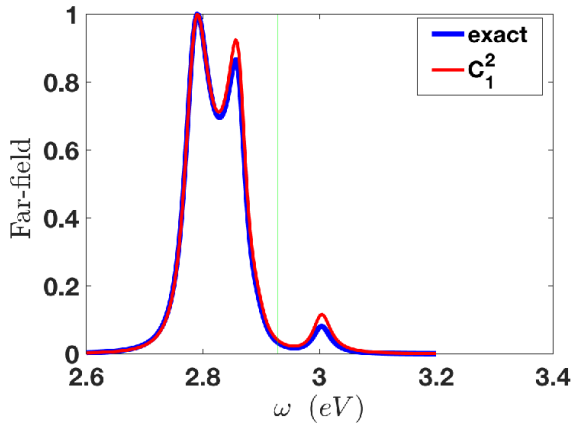}
\caption{Spectrum calculated at the detector position (Eq. (\ref{eq:Sfar}), 'exact') or from LSP$_1$ population $\vert C_1(\omega) \vert^2$. The vertical line refers to the TLS emission frequency. } 
\label{fig:FarField}
\end{figure}

However, the polarization spectrum lacks information on radiative and non radiative emission processes. Moreover, experimental characterization generally relies on far-field emission spectrum. Qualitative understanding of the far-field behaviour can be achieved considering the dipolar LSP$_1$ mode population ($ \vert C_1(\omega)  \vert ^2$). Indeed,  the far-field radiated signal (in the whole space)  can be written as
\begin{eqnarray}
P_{rad}&=&\frac{1}{2\pi}\gamma^{rad} P(\omega) \sim \vert C_1(\omega)  \vert ^2
\end{eqnarray}
where $\gamma^{rad}$ is the radiative decay rate (at the angular frequency $\omega$) in presence of the MNP (see \ref{sect:FarFieldSPec} for the details). We compare in Fig. \ref{fig:FarField} the far-field emission at the detector position and  the bright dipolar mode population spectrum. We obtain very good agreement, justifying that far-field emission is governed by the dipolar LSP$_1$ mode scattering. We observe again a Rabi splitting of $144$ meV that is a reminiscence of the strong coupling regime observed in the polarization spectrum. However, the main contribution comes from the bright LSP$_1$ scattering near $\omega=2.79$ eV. 

\section{Weak coupling regime: Purcell factor and Fermi's golden rule}
\label{sect:weak}
We are particularly interested in the role of losses on the hybrid system dynamics. In the previous section, we have discussed the strong coupling regime and the effect of both absorption and radiative losses on the near-field and far-field spectra. Before investigating in detail the effective Hamiltonian modification in presence of important radiative losses, it is worthwhile to discuss the dynamics of the coupled system in the weak coupling regime. This permits to introduce notably the Purcell factor before discussing how it is modified and why introducing Fano states in presence of radiative leakages. 

From the effective Hamiltonian (eq. \ref{eff_hamil}), the emitter and LSP population dynamics obeys 
\begin{eqnarray}
\label{eq:dCe}
\dot C_e(t)&=&-\frac{\gamma_0}{2} C_e(t) -i\sum_{n=1}^{N}g_nC_n(t) \;, \\
\dot C_n(t)&=&-ig_nC_e(t)-\left(i\Delta_n+\frac{\Gamma_n}{2}\right)C_n(t) \,.
\label{eq:dCn}
\end{eqnarray}
In the weak coupling regime, the population of the plasmon states remain small so that they can be adiabatically eliminated, that is $\dot C_n(t)\approx 0$. We obtain
 \begin{eqnarray}
 \nonumber
\dot C_e(t)&=&- \left[\frac{\gamma_0}{2}C_e(t)+\sum_{n=1}^{N}\frac{g_n^2}{i\Delta_n+\Gamma_n/2} \right] C_e(t) \;, \\
C_e(t)&=&C_e(0)e^{-i\delta \omega t} e^{-\frac{\gamma_{tot}}{2}t} \;,
\label{eq:CeWeak}
\end{eqnarray}
where we recognize the Lamb shift and the total decay rate
\begin{eqnarray}
\label{eq:Lamb}
\delta \omega&=&\sum_{n=1}^{N}\frac{g_n^2(\omega_0-\omega_n)}{(\omega_0-\omega_n)^2+(\Gamma_n/2)^2} \;, \\
\gamma_{tot}&=&\gamma_0+\sum_{n=1}^{N}\frac{g_n^2\Gamma_n}{(\omega_0-\omega_n)^2+(\Gamma_n/2)^2} \,.
\label{eq:Gtot}
\end{eqnarray}
We can define the Purcell factor $F_p^n$  for each mode such that 
\begin{eqnarray}
\label{eq:Fpn}
\frac{\gamma_{tot}}{\gamma_0}&=&1+\sum_{n=1}^{N}F_p^n \frac{1}{1+4Q_n^2\left(\frac{\omega_0-\omega_n}{\omega_n}\right)^2} \\
F_p^n &=&\frac{4g_n^2}{\gamma_0\Gamma_n}
\label{eq:Purcell0}
\end{eqnarray}
where $Q_n=\omega_n/\Gamma_n$ is the LSP$_n$ quality factor. Note that inserting the definition of the coupling strength (Eqs. \ref{lien} and \ref{structuration}) into the total decay rate (Eq. \ref{eq:Gtot}), we recover the Fermi golden rule result \cite{GCF-Barthes-Girard:2016}
\begin{eqnarray}
\nonumber
\frac{\gamma_{tot}}{\gamma_0}&=&1+\frac{2 k_0^2}{\hbar \epsilon_0 \gamma_0}\sum_{n=1}^{N} Im\left[ \mathbf{d}_{eg}\cdot\mathbf{G}_{\omega_0,n}(\mathbf{r}_d,\mathbf{r}_d) \cdot \mathbf{d}_{eg}^\star \right]  \\
&=&1+\eta \frac{6\pi}{k_b}Im G^{uu}_{\omega_0,n}({\bf r_d},{\bf r_d}) 
\label{eq:Fermi}
\end{eqnarray}
where $\eta$ is the intrinsic quantum yield $\eta=\gamma_0^{rad}/\gamma_0$ and $k_b=n_b k_0$. $G^{uu}_{\omega_0,n}=\mathbf{u}\cdot \mathbf{G}_{\omega_0,n}\cdot \mathbf{u}$ with $\mathbf{u}$ an unitary vector  along the TLS dipole moment ($\mathbf{d}_{eg}={d}_{eg}\mathbf{u}$).
Similarly, if we assume that the Green tensor follows a first order resonance (see {\it e.g.} Eq. \ref{eq:Gorder1} in \ref{sect:AnnexRate}), the Lamb shift (Eq. \ref{eq:Lamb}) can be rewritten as
\begin{eqnarray}
\frac{\delta\omega}{\gamma_0}&=&-\eta \frac{3\pi}{k_b}Re G^{uu}_{\omega_0,n}({\bf r_d},{\bf r_d}) 
\label{eq:Lamb}
\end{eqnarray}
This Lamb shift is generally negligible and will not be considered in the following. The normalized Fermi golden rule (Eq. \ref{eq:Fermi}) and Lamb shift (Eq. \ref{eq:Lamb}) are  in full agreement with the result obtained from classical approach considering an oscillating dipole \cite{Metiu:1984}.

\begin{figure}[h]
\includegraphics[width=8cm]{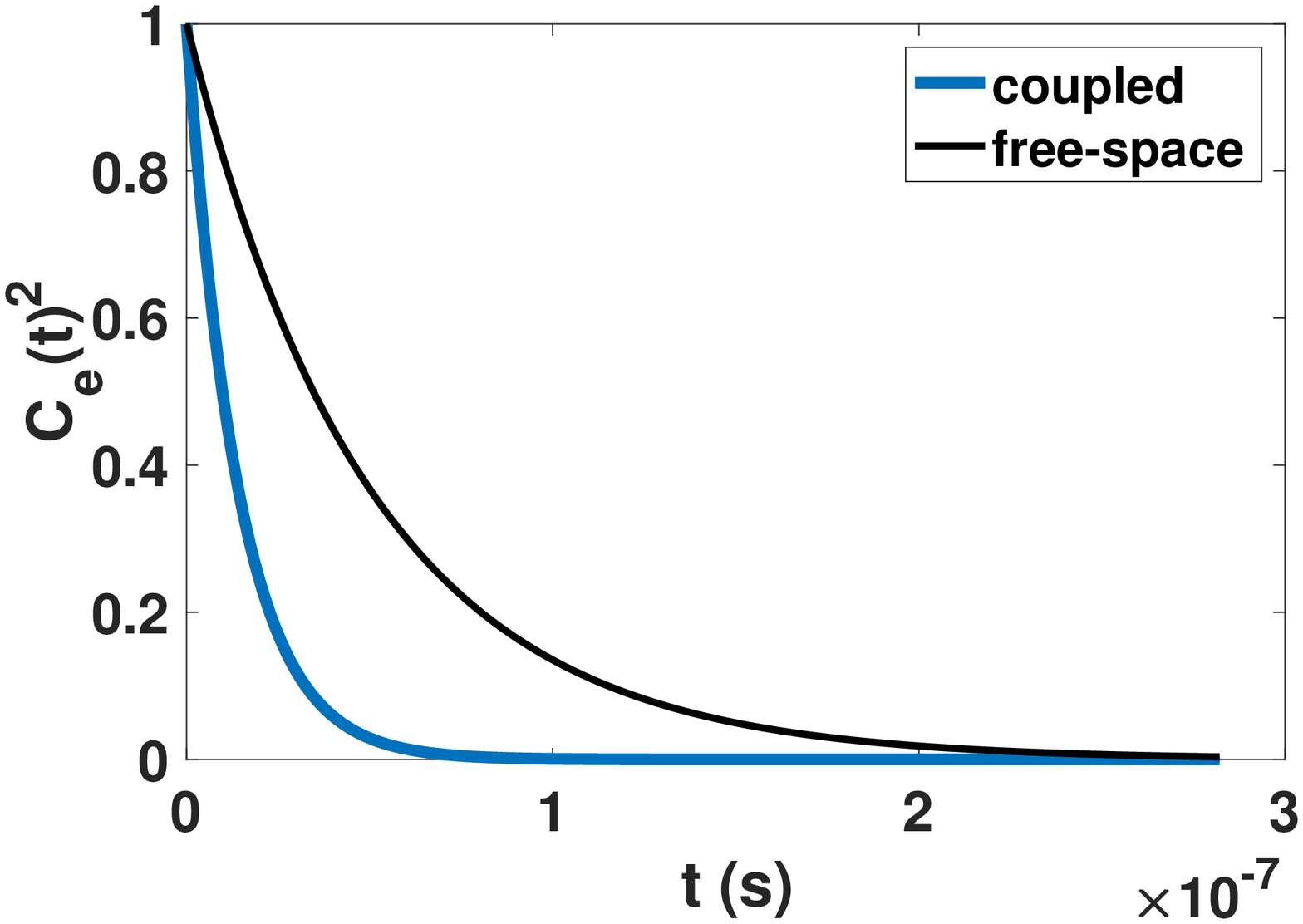}
\includegraphics[width=8cm]{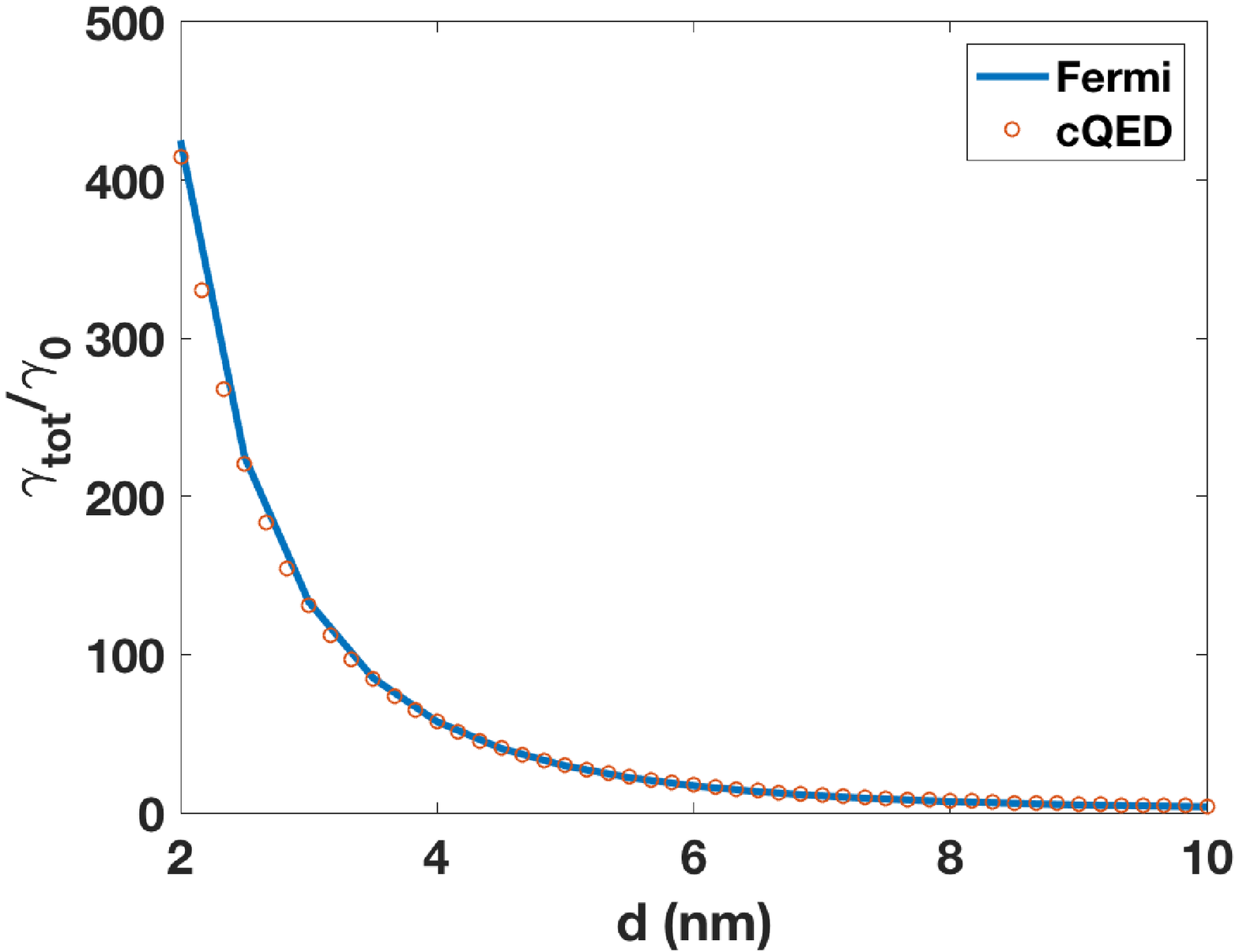}
	\caption{a) Excited state dynamics obtained from the effective Hamiltonian (Eq. \ref {eq:PsiDyn}) for a TLS 5 nm away from a silver nanoparticle (radius $R=8$ nm).  b) Normalized decay rate as a function of the distance to the silver nanoparticle calculated within the adiabatic elimination approximation (Eq. \ref{eq:Fpn} - dots)  or  Fermi's golden rule (Eq. \ref{eq:Fermi}- solid line).}
\label{fig:Decay}
\end{figure}

We consider in Fig. \ref{fig:Decay} a typical TLS emitting at $\lambda=670$ nm with $\tau_0=1/\gamma_0=50$ ns and presenting an intrinsic quantum yield $\eta=90 \%$. This corresponds to a dipole transition moment $d_{eg}=3.4$ D. The dynamics of the excited state close to the MNP is described in Fig. \ref{fig:Decay}a). In this weak coupling regime,  we observe an exponential decay with a fluorescence lifetime $\tau=1.7$ ns in agreement with Fermi's golden rule ($\gamma_{tot}/\gamma_0=\tau_0/\tau=30$ at d=5 nm, see Fig. \ref{fig:Decay}b).

Actually, the TLS emission is not limited to a single value $\omega_0$ but rather follows a Lorentzian profile. The contribution to the decay rate originating from the coupling to the LSP$_n$ is therefore 
\begin{eqnarray}
\gamma_n=\int_{-\infty}^\infty d\omega {\cal L}(\omega)\frac{g_n^2\Gamma_n}{(\omega-\omega_n)^2+(\Gamma_n/2)^2} \;, \\
{\cal L}(\omega)=\frac{\gamma_0/2\pi}{(\omega-\omega_0)^2+(\gamma_0/2)^2} 
\nonumber
\end{eqnarray}
where ${\cal L}(\omega)$ is the free-space normalized emission spectrum of the TLS. This is the analogue to the cQED description. Following the work of van Exter and coworkers \cite{vanExter-Woerdman:1996}, we can solve the integration over $\omega$ as $\gamma_n=2\pi g_n ^2 C_{{\cal L}{\cal L}_n}(0)$ where $C_{{\cal L}{\cal L}_n}(u)=\int_{-\infty}^\infty {\cal L}(\omega) {\cal L}_n(u-\omega)d\omega$ is the convolution product with a normalized  Lorentzian profile peaked at $-\omega_n$ with a FWHM $\Gamma_n$. The convolution of two normalized Lorentzians is also a normalized Lorentzian and it yields
\begin{eqnarray}
\gamma_n&=&\frac{g_n^2(\gamma_0+\Gamma_n)}{(\omega_0-\omega_n)^2+\left[(\gamma_0+\Gamma_n)/2\right]^2} \,.
\label{eq:GammaAmbiant}
\end{eqnarray}

Even at ambient temperature, $\gamma_0 \ll \Gamma_n$ so that expressions (\ref{eq:Gtot},\ref{eq:Fermi}) remain valid without working at cryogenic temperatures as for cQED. As expected, because of the strong subwavelength mode confinement, quantum plasmonics permits to transpose cQED behaviour to ambient temperature \cite{Baumberg:2016}.

\section{Leaky modes and LSP Fano states}
\label{sect:Fano}
\begin{figure}[h!]
\includegraphics[width=10cm]{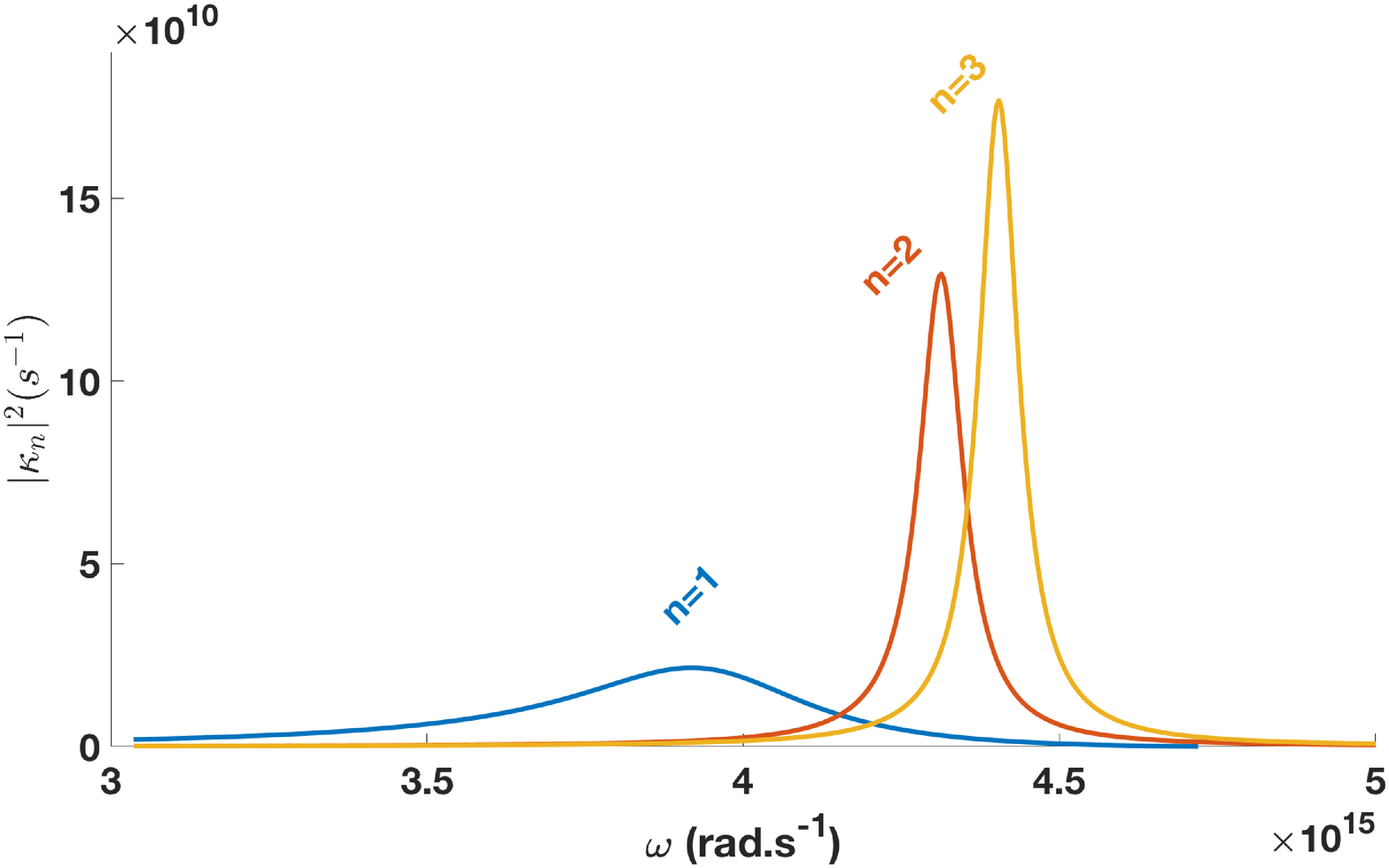}
\includegraphics[width=7cm]{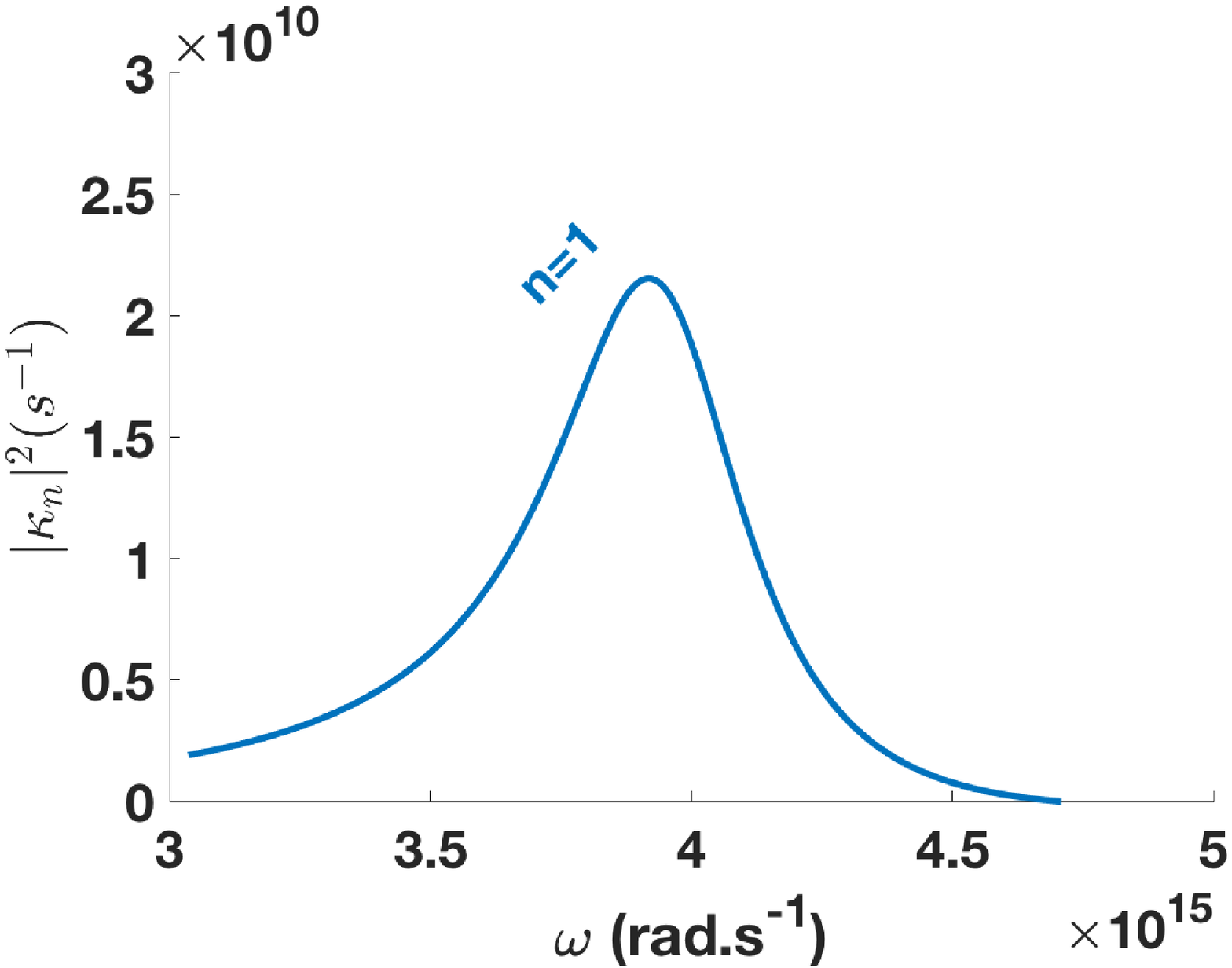}
\caption{Coupling constant spectra to LSP$_n$ modes calculated using Eq. (\ref{lien}). The silver particle radius is 50 nm and the emitter is located at 5nm from its surface. The right frame is an enlargement for $n=1$ showing the asymmetry.}
\label{fig:FanoMNP}
\end{figure}
For large MNP, the LSP modes can become strongly leaky, deforming the coupling strength spectrum, as shown in Fig. \ref{fig:FanoMNP}. Indeed, the dipolar plasmon LSP$_1$ becomes strongly radiative so that $\vert \kappa_1 \vert ^2$  does not follow a Lorentzian profile anymore \cite{GCF-Barthes-Girard:2016}. In the following, we discuss how to modify the effective Hamiltonian to include this behaviour. At this point, it is necessary to recall that the contribution of the free-space contribution was phenomenologically introduced in the first section (see the discussion on Eq. (\ref{eq:OpE})). In order to improve the effective non hermitian Hamiltonian, we consider the bath model inferred from the effective Hamiltonian (Fig. \ref{fig:BathcQED}) as a starting point. In order to clarify the role of the free space contribution, we first focus on lossless TLS and MNP in section \ref{sect:AnnexFano} before proposing a general effective Hamiltonian in section \ref{sect:FanoEff}.
\begin{figure}[h!]
\includegraphics[width=16cm]{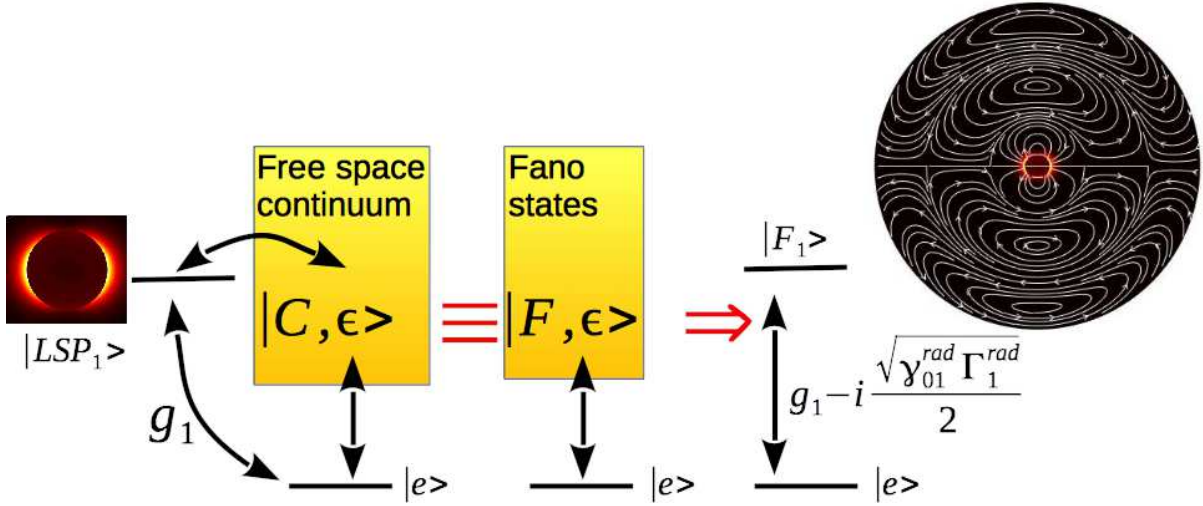}
\caption{Open emitter-LSP$_1$ quantum system in absence of absorption. Left) The excited state of the emitter and the LSP$_1$ state are coupled to the same reservoir associated to radiation leakages. The inset shows the mode profile. Middle) Equivalent open quantum system after Fano diagonalization. Right) Corresponding effective Hamiltonian resulting from the coupling to the leaky Fano dipolar plasmon $\vert F_1 \rangle$.  The inset shows the mode near-field intensity as well as the electric field lines revealing the far-field leakages. The particle radius is 50 nm and the windows size is 700 nm (the resonance wavelength is $\lambda_1=2\pi c/\omega_1=477$ nm. }
\label{fig:FanoBathcQED}
\end{figure} 

\subsection{Fano states for lossless TLS and MNP}
\label{sect:AnnexFano}
\subsubsection{Heuristic presentation of Fano states.}
Let us consider the ideal situation without absorption (TLS intrinsic quantum yield $\eta=100\%$ and lossless metal $\Gamma_p=0$). Radiation into the far-field is the only available channel for energy dissipation. Moreover, all LSPs modes are dark, except LSP$_1$ dipolar mode, for not too large particles. Therefore,  we consider the open quantum system schemed in Fig. \ref{fig:FanoBathcQED} where the emitter and LSP$_1$ are coupled to the same radiation bath and we restrict the discussion to a single mode (LSP$_1$) MNP for the sake of clarity.
In analogy to the  work of Knight {\it et al} \cite{Knight:90,Durand-Gadea:01} we define a Fano state $\vert F \rangle$ diagonalizing the LSP$_1$-bath states and construct a modified effective Hamiltonian (see also \S C$_I$ in ref. \cite{Cohen-Tannoudji1996}) in the basis $\{\vert e,\vac\rangle,\vert g, F\rangle \}$ 
\begin{eqnarray}
H_{eff}=\hbar \left[
\begin{array}{cc}
\omega_0 -i\frac{\gamma_0^{rad}}{2} & g_1-\frac{i}{2}\sqrt{\gamma_0 ^{rad}\Gamma_1^{rad} }\\
g_1-\frac{i}{2}\sqrt{\gamma_0^{rad} \Gamma_1^{rad}}  &\omega_1 -i\frac{\Gamma_1^{rad}}{2} 
 \end{array}
\right] \,.
\label{eq:HFano2}
\end{eqnarray}
$\gamma_0^{rad}$ and $\Gamma_1^{rad}$ define the radiative contribution to the decay rates since no absorption is considered here.  
\subsubsection{Microscopic derivation of LPs Fano states.}
Generalization of the effective Hamiltonian (\ref{eq:HFano2}) to all LSPs is not straightforward since the emitter and all LSPs could couple to the free-space continuum for large particle. Therefore, we go back the bath model inferred in Fig. \ref{fig:BathcQED} where each LSP mode is associated to a specific reservoir. We add the direct emitter-radiative bath coupling ($\zeta_n$) so that we consider the following Hamiltonian (compare with Eq. \ref{H_SE})
\begin{eqnarray}
\label{H_SEFano}
\hat{H'}_{SE}&=&H_S+H_E+H'_I,\\
\nonumber
H'_{I}&=&i\hbar\int d\omega\,\sum_{n=1}^\infty\beta_n^r(\omega)\left(\hat{b}_{\omega,n}^{r \dagger}\hat{a}_n-\hat{a}_{n}^{\dagger}\hat{b}_{\omega,n}^r \right)
 \\
\nonumber
&&+i\hbar\int d\omega\,\sum_{n=1}^\infty\zeta_n(\omega)\left(\hat{b}_{\omega,n}^{r \dagger}\hat{\sigma}_{ge}-\hat{\sigma}_{eg}^{\dagger}\hat{b}_{\omega,n}^r\right).
\end{eqnarray}
For flat couplings $\beta_n^r(\omega)\approx\sqrt{\Gamma_n^{rad}/2\pi}$, and $\zeta_n(\omega)\approx \sqrt{\gamma_{0n}^{rad}(\omega_0)/2\pi}$ it leads to the following Lindblad equation 
\begin{eqnarray}
\frac{d\hat{\rho}(t)}{dt}=\frac{1}{i\hbar} \left[\hat{H}_S,\hat{\rho}(t)\right]+\mathcal{D}_F\left[\hat{\rho}(t)\right] 
\label{LindbladFano}
\end{eqnarray}
where the new dissipator is
\begin{eqnarray}
\label{DissipatorF}
&&\mathcal{D}_F\left[\hat{\rho}(t)\right]=
-\frac{1}{2}\sum_{n=1}^\infty
\left[\hat{c}^{\dagger}_{n}\hat{c}_{n}\hat{\rho}(t)+\hat{\rho}(t)\,\hat{c}^{\dagger}_{n}\hat{c}_{n}-2\hat{c}_{n}\hat{\rho}(t)\,\hat{c}^{\dagger}_{n}\right] \;, \\
&&\hat{c}_{n}=\sqrt{\gamma_{0n}^{rad}(\omega_0)}\hat{\sigma}_{ge}+\sqrt{\Gamma_n^{rad}}\hat{a}_{n}\,.
\nonumber
\end{eqnarray}
$\gamma_{0n}^{rad}$ refers to the emitter relaxation into the radiation bath $\hat{b}_n$. The total radiative decay rate of the emitter in free-space obeys 
\begin{eqnarray}
\gamma_{0}^{rad}=\frac{2 k_0^2}{\hbar \epsilon_0}Im\left [ \mathbf{d}_{eg}\cdot \mathbf{G}_0(\mathbf{r}_d,\mathbf{r}_d,\omega_0) \cdot \mathbf{d}_{eg}^\star \right]
=n_b\frac{d_{eg}^2\omega_0^3}{3\pi\epsilon_0 \hbar c^3} 
\end{eqnarray}
so that $\gamma_{0n}^{rad}$ corresponds to the decomposition of this decay rate on the spherical harmonics and $\gamma_0^{rad}=\sum_{n=1}^\infty \gamma_{0n}^{rad}(\omega_0)$ (see \ref{sect:MieExpansion}).

Finally, the dissipator (Eq. \ref{DissipatorF}) can be written as the sum of three contributions: 
\begin{eqnarray}
\label{DissipatorS}
&&\mathcal{D}_F\left[\hat{\rho}(t)\right]=\mathcal{D}_0\left[\hat{\rho}(t)\right]+\sum_{n=1}^N \mathcal{D}_{LSP_n}\left[\hat{\rho}(t)\right]+\sum_{n=1}^N \mathcal{D}_{0,LSP_n}\left[\hat{\rho}(t)\right] \\
\nonumber
&&\mathcal{D}_0=-\frac{\gamma_0^{rad}}{2}\left[\hat{\sigma}_{eg}\hat{\sigma}_{ge}\hat{\rho}(t)+\hat{\rho}(t)\,\hat{\sigma}_{eg}\hat{\sigma}_{ge}-2\hat{\sigma}_{ge}\hat{\rho}(t)\hat{\sigma}_{eg}\right] \\
\nonumber
&&\mathcal{D}_{LSP_n}\left[\hat{\rho}(t)\right]=-\frac{\Gamma_n^{rad}}{2}\left[\hat{a}^{\dagger}_{n}\hat{a}_{n}\hat{\rho}(t)+\hat{\rho}(t)\,\hat{a}^{\dagger}_{n}\hat{a}_{n}-2\hat{a}_{n} \hat{\rho}(t)\,\hat{a}^{\dagger}_{n}\right] \; \\
\nonumber
&&\mathcal{D}_{0,LSP_n}\left[\hat{\rho}(t)\right]=-\frac{\sqrt{\gamma_{0n}^{rad}(\omega_0)\Gamma_n^{rad}}}{2}\left[\hat{\sigma}_{eg}\hat{a}_n\hat{\rho}(t)+\hat{\rho}(t)\,\hat{\sigma}_{eg}\hat{a}_n-2\hat{a}_{n}\hat{\rho}(t)\hat{\sigma}_{eg}\right. \\
&& \hspace{6cm}\left. +\hat{a}_n^\dagger\hat{\sigma}_{ge}\hat{\rho}(t)+\hat{\rho}(t)\,\hat{a}_n^\dagger\hat{\sigma}_{ge}-2\hat{\sigma}_{ge}\hat{\rho}(t)\hat{a}^{\dagger}_{n} 
\right] 
\nonumber
\end{eqnarray}
$\mathcal{D}_0$ and $\mathcal{D}_{LSP_n}$ describe the emitter and LSP$_n$ relaxations, respectively and $\mathcal{D}_0$ naturally appears without the need of phenomenological introduction (see also Eq. \ref{Lindblad}). $\mathcal{D}_{0,LSP_n}$ refers to a collective relaxation process that originates from their coupling to the same bath. Let us note that a similar collective dissipator has been recently phenomenologically introduced to interprate enhanced optical trapping of an assembly of emitters \cite{JuanPRL:18}.

Following the work of Visser and Nienhuis \cite{Visser-Nienhuis:95}, it is straightforward to build an effective Hamiltonian from the Lindblad master equation (\ref{LindbladFano}). It obeys 
\begin{eqnarray}
H_{eff}=\hat{H}_S-i\sum_{n=1}^N \frac{1}{2}\hat{c}^{\dagger}_{n}\hat{c}_{n}
\end{eqnarray}
so that we obtain 
\begin{eqnarray}
\label{HeffFano}
\hspace{-2cm}
&&H_{eff}=\hbar
\left(
\begin{array}{ccccc}
-i\frac{\gamma_0^{rad}}{2} & g_1-\frac{i}{2}\sqrt{\gamma_{01}^{rad}\Gamma_1^{rad}}& \cdots & g_N-\frac{i}{2}\sqrt{\gamma_{0N}^{rad}\Gamma_N^{rad}}\\
g_1-\frac{i}{2}\sqrt{\gamma_{01}^{rad}\Gamma_1^{rad}} & \Delta_1-i\frac{\Gamma_1^{rad}}{2} & 0 & \cdots \\
\vdots & \vdots & \ddots & \ddots \\
g_N -\frac{i}{2}\sqrt{\gamma_{0N}^{rad}\Gamma_N^{rad}}&  \cdots & 0 & \Delta_N-i\frac{\Gamma_N^{rad}}{2}
\end{array}
\right)\; 
\end{eqnarray}

\subsection{General non hermitian effective Hamiltonian}
\label{sect:FanoEff}
\begin{figure}
\includegraphics[width=15cm]{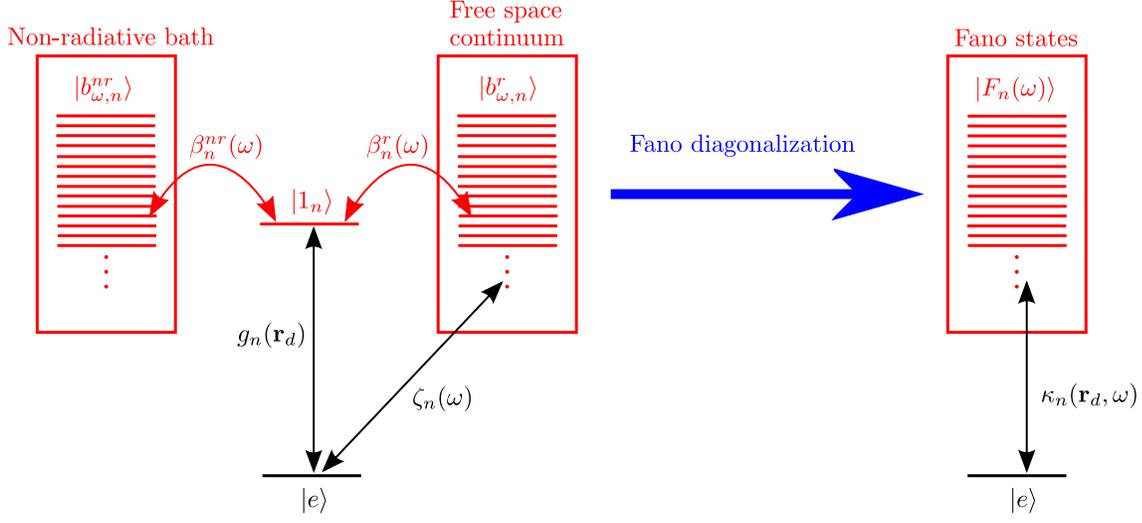}
\caption{Open emitter-LSPs quantum system considering metal absorption. The emitter and the LSPs couple to a radiative bath. LSPs are also couple to a non- radiative bath. Right) Equivalent open quantum system after Fano diagonalization.}
\label{fig:FanoNR}
\end{figure} 
We now take into account the absorption into the metal ($\Gamma_p \ne0 $) but still consider TLS intrinsic quantum yield $\eta=100\%$. The LSP$_n$ couple to an additionnal non radiative bath as schemed in Fig. \ref{fig:FanoNR}. The Hamiltonian is written as
\begin{eqnarray}
\label{H_SEFano}
\hat{H'}_{SE}&=&H_S+H'_E+H'_I,\\
\nonumber
H'_E &=&\int d\omega\,\hbar\omega\sum_{n=1}^\infty\hat{b}_{\omega,n}^{r \dagger}\hat{b}_{\omega,n}^r 
+\int d\omega\,\hbar\omega\sum_{n=1}^\infty\hat{b}_{\omega,n}^{nr \dagger}\hat{b}_{\omega,n}^{nr} ,\\
\nonumber
H'_{I}&=&i\hbar\int d\omega\,\sum_{n=1}^\infty \beta_n^r(\omega)\left(\hat{b}_{\omega,n}^{r \dagger}\hat{a}_n-\hat{a}_{n}^{\dagger}\hat{b}_{\omega,n}^r\right)
 \\
\nonumber
&&+i\hbar\int d\omega\,\sum_{n=1}^\infty\zeta_n(\omega)\left(\hat{b}_{\omega,n}^{r \dagger}\hat{\sigma}_{ge}-\hat{\sigma}_{eg}^{\dagger}\hat{b}_{\omega,n}^r\right) 
 \\
\nonumber
&&+i\hbar\int d\omega\,\sum_{n=1}^\infty\beta_n^{nr}(\omega)\left(\hat{b}_{\omega,n}^{nr \dagger}\hat{a}_n-\hat{a}_n^{\dagger}\hat{b}^{nr}_{\omega,n}\right).
\end{eqnarray}
This leads to the following Lindblad equation 
\begin{eqnarray}
\frac{d\hat{\rho}(t)}{dt}=\frac{1}{i\hbar} \left[\hat{H}_S,\hat{\rho}(t)\right]+\mathcal{D}_F\left[\hat{\rho}(t)\right] +\mathcal{D}_{nr}\left[\hat{\rho}(t)\right] 
\label{LindbladFano}
\end{eqnarray}
whith the additional non-radiative dissipator
\begin{eqnarray}
\label{DissipatorNR}
&&\mathcal{D}_{nr}\left[\hat{\rho}(t)\right]=
-\frac{1}{2}\sum_{n=1}^N \Gamma_n^{nr}
\left[\hat{a}^{\dagger}_{n}\hat{a}_{n}\hat{\rho}(t)+\hat{\rho}(t)\,\hat{a}^{\dagger}_{n}\hat{a}_{n}-2\hat{a}_{n}\hat{\rho}(t)\,\hat{a}^{\dagger}_{n}\right]\,.
\nonumber
\end{eqnarray}
leading to the effective Hamiltonian 
\begin{eqnarray}
\label{HeffFano}
\hspace{-2cm}
&&H_{eff}=\hbar
\left(
\begin{array}{ccccc}
-i\frac{\gamma_0^{rad}}{2} & g_1-\frac{i}{2}\sqrt{\gamma_{01}^{rad}\Gamma_1^{rad}}& \cdots & g_N-\frac{i}{2}\sqrt{\gamma_{0N}^{rad}\Gamma_N^{rad}}\\
g_1-\frac{i}{2}\sqrt{\gamma_{01}^{rad}\Gamma_1^{rad}} & \Delta_1-i\frac{\Gamma_1}{2} & 0 & \cdots \\
\vdots & \vdots & \ddots & \ddots \\
g_N -\frac{i}{2}\sqrt{\gamma_{0N}^{rad}\Gamma_N^{rad}}&  \cdots & 0 & \Delta_N-i\frac{\Gamma_N}{2}
\end{array}
\right)\; 
\end{eqnarray}
with $\Gamma_n=\Gamma_n^{rad}+\Gamma_n^{nr}$ the total decay rate of the LSP$_n$. We observe that the non radiative processes increases the LSPs losses but does not play a role on the emitter-LSP coupling (off-diagonal elements). Actually, the Lindblad equation (Eq. \ref{Lindblad}) proposed for small (non leaky but absorbing) MNP corresponds to LSPs couple to the non-radiative bath only but the  emitter coupled to the radiative bath,  explaining why the emitter spontaneous emission ($\mathcal{D}_0$) was phenomenologically introduced whereas it naturally appears within the bath model of Fig. \ref{fig:FanoNR}.

Finally, the effective Hamiltonian formally takes the form in the LSPs Fano state basis $\{\vert e,\vac\rangle,\vert g, F_1\rangle,\cdots,\vert g, F_N\rangle \}$
\begin{eqnarray}
\label{HeffFano}
\hspace{-2cm}
H_{eff}=\hbar 
\left(
\begin{array}{ccccc}
-i\frac{\gamma_0}{2} & g_1[1-\frac{i}{2}\alpha_1(\omega_0)] &  g_2[1-\frac{i}{2}\alpha_2(\omega_0)] & \cdots & g_N[1-\frac{i}{2}\alpha_N(\omega_0)]\\
g_1[1-\frac{i}{2}\alpha_1(\omega_0)]& \Delta_1-i\frac{\Gamma_1}{2} & 0 & \cdots & 0\\
g_2[1-\frac{i}{2}\alpha_2(\omega_0)] & 0 & \Delta_2-i\frac{\Gamma_2}{2} & \ddots & \vdots\\
\vdots & \vdots & \ddots & \ddots & 0\\
g_N[1-\frac{i}{2}\alpha_N(\omega_0)] & 0 & \cdots & 0 & \Delta_N-i\frac{\Gamma_N}{2}
\end{array}
\right)\; \hspace{.4cm}
\end{eqnarray}
where $\alpha_n$ is the ratio between the coupling to the radiative quasi-continuum ($\alpha_n(\omega_0)g_n=\sqrt{\gamma_{0n}^{rad}(\omega_0)\Gamma_n^{rad}}$) and the coupling to the discrete LSP$_n$ state (given by $g_n$). It depends on the emission angular frequency $\omega_0$. $q_{F,n}=2/\alpha_n$ is the Fano parameter for the $n^{th}$ mode. It is equivalent to the exact discrete form (Eq. \ref{eff_hamil}) in the limit $\alpha_n \rightarrow 0$ corresponding to a negligible coupling to the quasi-continuum. 
\subsection{Weak coupling regime}
We now discuss how the excited emitter dynamics is modified considering the new effective Hamiltonian. Here again, we consider adiabatic elimination in the weak coupling regime. The population dynamics follows then
 \begin{eqnarray}
C_e(t)&=&C_e(0)e^{-i\delta \omega t} e^{-\frac{\gamma_{tot}}{2}t}
\end{eqnarray}
where
\begin{eqnarray}
\label{eq:GtotFano}
\delta \omega&=&-\sum_{n=1}^{N}\frac{g_n^2}{\Delta_n^2+(\Gamma_n/2)^2} \left[(1-\frac{\alpha_n^2}{4})\Delta_n +\alpha_n\frac{\Gamma_n}{2}\right]\;, \\
\gamma_{tot}&=&\gamma_0+\sum_{n=1}^{N}\frac{g_n^2}{\Delta_n^2+(\Gamma_n/2)^2}\left[(1-\frac{\alpha_n^2}{4})\Gamma_n-2\alpha_n\Delta_n\right] \;.
\nonumber
\end{eqnarray}
We can again express the total decay rate such that 
\begin{eqnarray}
\frac{\gamma_{tot}}{\gamma_0}&=&1+\sum_{n=1}^{N}\frac{\gamma_n}{\gamma_0} \;, \\
\label{eq:FpnFano}
\frac{\gamma_n}{\gamma_0} &=&F_p^n(\omega_0) \frac{1}{1+4Q_n^2\left(\frac{\omega_0-\omega_n}{\omega_n}\right)^2} \left[1-\frac{\alpha_n^2(\omega_0)}{4}+2\alpha_n(\omega_0)Q_n\frac{\omega_0-\omega_n}{\omega_n}\right] \\
\nonumber
\end{eqnarray}
$\gamma_n$ refers to the decay rate into LSP$_n$. The dependency of the parameters on the emission frequency $\omega_0$ is explicitely indicated.

Since $\alpha_n$ only slightly depends on the emission frequency $\omega_0$ around $\omega_n$, the Fano shape of the decay rate (Eq. \ref{eq:FpnFano}) is similar (at the first order in $\alpha_n$) to the one obtained by Sauvan and coworkers \cite{Sauvan-Lalanne:2013} considering a fully classical treatment of the atom-leaky mode coupling. They generalized the effective volume of the mode by defining a complex mode volume $\widetilde{V_n}$ such that the Fano parameter obeys $2/q_{F,n}\approx \alpha_n(\omega_n)=Im(\widetilde{V_n})/Re(\widetilde{V_n})$, that characterizes the TLS coupling branch ratio to  the leaky (quasi-continuum) and the discrete contribution to LSP$_n$ modes. Specifically, they use quasi-normal mode analysis that accurately describes the mode leakage. Since QNM corresponds to the pole of the Green tensor, the two approaches are fully equivalent, making a bridge between cQED and classical approaches in the weak coupling regime.

\begin{figure}
\includegraphics[width=8cm]{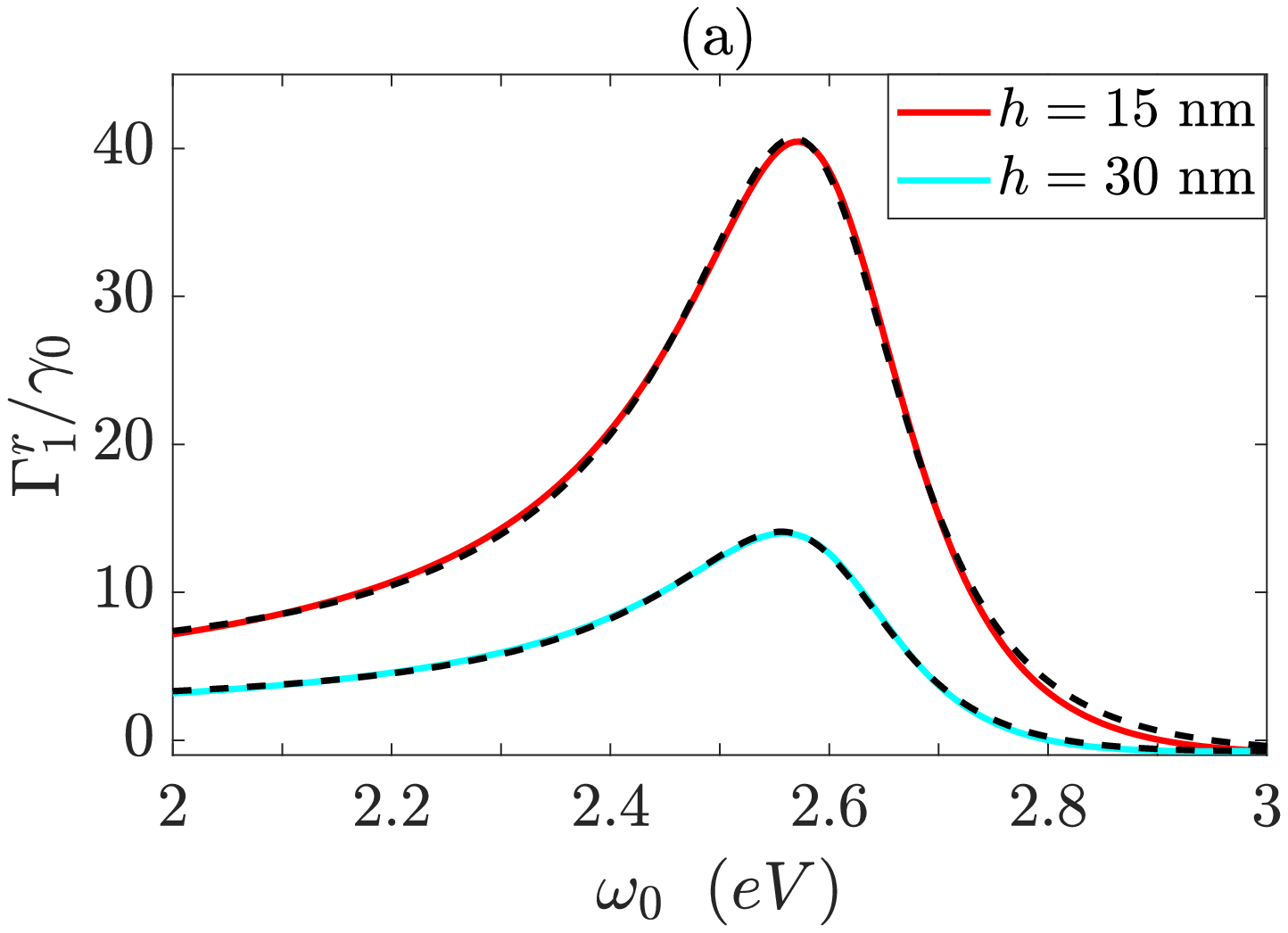}
\includegraphics[width=8cm]{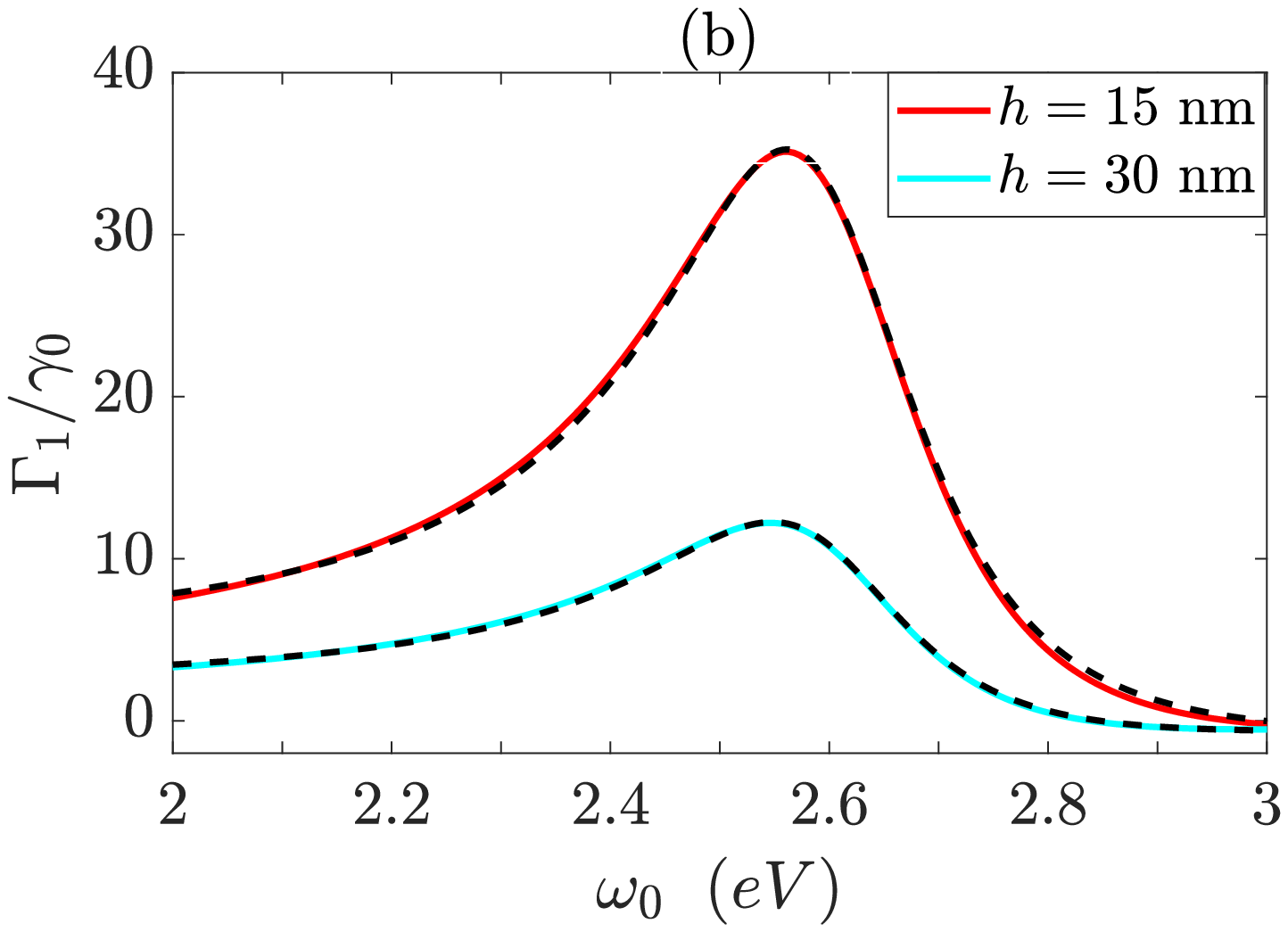}
\caption{Normalized decay rate into the LSP$_1$ dipolar mode as a function of the emission wavelength. a) Non absorbing systems ($\eta=100 \%$ for the TLS and $\hbar \Gamma_p=0$ for the MNP). Dots refer to Fano fit following Eq. (\ref{eq:FpnFano}) with $\omega_1=2.60$ meV, $\Gamma_1^{rad}=254$ meV and $g_1=-6.91\times 10^{-2}$  meV ($h=15$ nm) or $g_1=-4.03 \times10^{-2}$  meV ($h=30$ nm). b) Absorbing MNP ($\hbar \Gamma_p=51$ meV). Dots refer to Fano fit following Eq. (\ref{eq:FpnFano}) with the additionnal parameter $\Gamma_1^{nr}=40$ meV. $d_{eg}=1$D.} 
\label{fig:AgFano}
\end{figure}

Assuming that the decay rate still obeys to the classical Fermi golden rule (Eq. \ref{eq:Fermi}), we plot in Fig. \ref{fig:AgFano}a) the decay rate to LSP$_1$ for a large (leaky) non absorbing silver nanoparticle. We observe a Fano behaviour in agreement with expression (\ref{eq:FpnFano}). There are only three fitting parameters, namely $\omega_1, \Gamma_1^{rad}$  and $g_1$. The Fano parameter obeys $q_F=2/\alpha_1(\omega_0)=2g_1/\sqrt{\gamma_{01}^{rad}(\omega_0)\Gamma_1^{rad}}=-4.2$ at $\omega_0=\omega_1$ for $h=30$ nm and the Purcell factor is obtained from $F_{rad}^1 =4g_1^2/\gamma_0^{rad}\Gamma_1^{rad}=14.2$ ($F_{rad}^1=40.7$ for $h=15$ nm). In Fig. \ref{fig:AgFano}b), we include metal absorption. We use the same $\omega_1, \Gamma_1^{rad}$  and $g_1$ as for lossless configuration and fit the decay rate spectral behaviour with one additionnal parameter; the LSP non-radiative rate $\Gamma_1^{nr}$.
Finally, one can write the Purcell factor in presence of the lossy MNP as
\begin{eqnarray}
F_p^n=\frac{4g_n^2}{\gamma_0^{rad}\Gamma_n}=\frac{\Gamma_n^{rad}}{\Gamma_n^{rad}+\Gamma_n^{nr}}F_{rad}^n \,.
\end{eqnarray}
The total Purcell factor is the LSP quantum yield times the lossless Purcell factor. LSP losses therefore decrease the Purcell factor independently of the distance (but the lossless Purcell factor depends on the distance to the MNP). We achieve $F_p^1=12.2$ for $h=30$ nm and $F_p^1=35.1$ for $h=15$ nm.

\section{Conclusion}
We have built effective Hamiltonians describing the emitter-MNP interaction extending the cQED approach to quantum plasmonics. We extensively discuss the role of Joule and radiative losses in the coupling process and their effect on the Hamiltonian structure. Because the effective Hamitonian of the hybrid nanosource is time independent, we can introduce true energy levels (so called dressed states) and discuss the effect of atom-plasmon interaction on the wavefunction of the coupled system. We also discussed the link between near-field and far field spectra with the population of the emitter and radiative dipolar plasmon, respectively. Moreover, with quantized plasmon field, we can clearly identify the elementary process of spontaneous emission and we define a Purcell factor for each LSP. For large particles, we observe a Fano profile, fully explained considering a modified effective Hamiltonian, inspired from cQED considerations. We also derive Lindblad equations for each situation and introduce a collective dissipator for describing the Fano behaviour. This clarify the role of radiative leakages (spontaneous emission) and overcome the difficulty of their phenomenological introduction that misses this collective dissipator. Finally, we stress that our formalism directly transposes cQED concepts to the nanoscale and constitutes therefore a powerful tool to propose and design ultrafast nanophotonics devices, taking benefit of the mode subwavelength confinement. 
\section{acknowledgements}
This work was supported by the French "Investissements d'Avenir" program, through the project ISITE-BFC IQUINS (contract ANR-15-IDEX-03 ) and EUR-EIPHI contract (17-EURE-0002). This work is part of the  european COST action MP1403 Nanoscale Quantum Optics.

\onecolumn
\appendix
\section{Spherical particle - Mie expansion of the Green dyad}
\label{sect:MieExpansion}
The Green dyad associated to the spherical particle is expressed using the Mie expansion \cite{Chew:1987,Hakami-Wang-Zubairy:2014}
\begin{eqnarray}
{\bf G}_S({\bf r},{\bf r_d})&=&\frac{ik_b}{4\pi}\sum_{e,o}\sum_{n=1}^\infty \sum_{m=0}^n (2-\delta_{0m})\frac{2n+1}{n(n+1)}\frac{(n-m)!}{(n+m)!} \\
\nonumber
&& \hspace{1cm}\left[A_n{\bf M}_{mn}^{(1)}({\bf r})\otimes {\bf M}_{mn}^{(1)}({\bf r_d})
+B_n{\bf N}_{mn}^{(1)}({\bf r})\otimes {\bf N}_{mn}^{(1)}({\bf r_d}) 
 \right] 
\end{eqnarray}
The formula of the spherical vector wave functions ${\bf M},{\bf N}$ can be found in ref. \cite{Hakami-Wang-Zubairy:2014}.
The Mie coefficients are
\begin{eqnarray}
A_n=\frac{j_n(k_mR)\psi_n'(k_bR)-j_n(k_bR)\psi_n'(k_mR)}{h_n{(1)}(k_bR)\psi_n'(k_mR)-j_n(k_mR)\zeta_n'(k_bR)} \,,\\
B_n=\frac{k_b^2j_n(k_bR)\psi_n'(k_mR)-k_m^2j_n(k_mR)\psi_n'(k_bR)}{k_m^2j_n(k_mR)\zeta_n'(k_bR)-k_b^2h_n^{(1)}(k_bR)\psi_n'(k_mR)}
\end{eqnarray}
where  $j_n$ and $h_n^{(1)}$ are the spherical Bessel and Hankel function. $\psi_n(z)=zj_n(z)$, $\zeta_n(z)=zh_n^{(1)}$ the Ricatti-Bessel functions. 
 
It can be also useful to expand the free-space Green tensor on the spherical harmonics. 
\begin{eqnarray}
\mathbf{G}_0(\mathbf{r},\mathbf{r_d})&=\frac{\delta(\mathbf{r}-\mathbf{r_d})}{k_b^2}\mathbf{e_r}\otimes\mathbf{e_r}+i\frac{k_b}{4\pi}\sum_{p=e,o}\sum_{n=0}^{+\infty}\sum_{m=0}^{n}(2-\delta_{0m})\frac{(2n+1)(n-m)!}{n(n+1)(n+m)!}\nonumber\\
&\times \left \{
\begin{array}{c}
\mathbf{M}^{(1)}_{nmp}\left(\mathbf{r}\right)\otimes\mathbf{M}^{(0)}_{nmp}\left(\mathbf{r_d}\right)+
\mathbf{N}^{(1)}_{nmp}\left(\mathbf{r}\right)\otimes\mathbf{N}^{(0)}_{nmp}\left(\mathbf{r_d}\right)\hspace{0.5cm}  r\geqslant r_d \label{G_0} \\
\\ \nonumber
\mathbf{M}^{(0)}_{nmp}\left(\mathbf{r}\right)\otimes\mathbf{M}^{(1)}_{nmp}\left(\mathbf{r_d}\right)+
\mathbf{N}^{(0)}_{nmp}\left(\mathbf{r}\right)\otimes\mathbf{N}^{(1)}_{nmp}\left(\mathbf{r_d} \right)\hspace{0.5cm} r\leqslant r_d
\end{array}
\right.
\end{eqnarray}
This expansion was used to calculate $\gamma_{0n}$ in \S \ref{sect:Fano}.

\section{Near-field coupling rate in the quasi-static approximation}
\label{sect:AnnexRate}
In the following, we consider a dipolar emitter with radial orientation since it corresponds to the most efficient coupling. 
Then, we get 
\begin{eqnarray}
G_S^{rr}({\bf r_d},{\bf r_d})=\frac{ik_b}{4\pi}\sum_{n=1}^\infty n(n+1)(2n+1) B_n\left[\frac{h_n^{(1)}(k_br_d)}{k_br_d} \right]^2
\label{eq:ImGzz}
\end{eqnarray}  
\subsection{Quasi-static approximation}
We assume that the sphere radius is very small compared to the wavelength, {\it i-e} $k_bR <<1 ,\vert k_mR\vert <<1$. Then, the Mie coefficient $B_n$ can be approximated to (with $u_b=k_bR$,$u_m=k_mR$)
\begin{eqnarray}
B_n
&\approx &
\frac{1}{i(2n-1)!!(2n+1)!!}
\frac{(n+1)u_b^n u_m^n(k_b^2-k_m^2)}{\left[n k_m^2+(n+1)k_b^2\right] u_m^n/u_b^{n+1}}\\
&\approx & k_b^{2n+1}
\frac{i(n+1)}{(2n-1)!!(2n+1)!!}
\frac{(\epsilon_S-\epsilon_B)R^{2n+1}}{n\epsilon_S+(n+1)\epsilon_B}
\end{eqnarray}
where we have used the limiting values \cite{Abramowitz-Stegun:1972}
\begin{eqnarray}
j_n(z)&\mathop{\sim}\limits_{z\rightarrow 0}\frac{z^n}{(2n+1)!!} \\
\psi_n'(z)&\mathop{\sim}\limits_{z\rightarrow 0} \frac{(n+1)z^n}{(2n+1)!!}\\
\label{eq:B_nQS}
h_n^{(1)}(z)&\mathop{\sim}\limits_{z\rightarrow 0}-i \frac{(2n-1)!!}{z^{n+1}}\\
\zeta_n'(z)&\mathop{\sim}\limits_{z\rightarrow 0} \frac{in(2n-1)!!}{z^{n+1}}
\end{eqnarray}

Finally, the Mie coefficient depends on the quasi-static polarisability for small particle size. Indeed, in the quasi-static regime,  the optical response of the particle can be described using a multipolar expansion.  If the particle excited with an incident field $ {\bf E_0}$, the $n^{th}$ multipole tensor moment is given by 
\begin{eqnarray}
{\bf p}^{(n)}&=&\frac {4 \pi \epsilon_0\epsilon_B} {(2n-1)!!} \alpha_n \nabla ^{n-1} {\bf E_0} \,,\\
\alpha_n&=&\frac{n(\epsilon_S-\epsilon_B)}{n\epsilon_S+(n+1)\epsilon_B}R^{2n+1}
\label{eq:alpha_n}
\end{eqnarray}
So that the approximate form of the Mie coefficient (Eq. \ref{eq:B_nQS}) can be rewritten as
\begin{eqnarray}
B_n&\approx&  i 
\frac{(n+1)k_b^{2n+1}}{n(2n-1)!!(2n+1)!!}\alpha_n
\label{eq:BnApprox}
\end{eqnarray}

\subsubsection{Resonance profile.}
For the sake of clarity, we assume that the surrounding medium is air, $\epsilon_b=1$ (see ref. \cite{GCF-Derom-Vincent-Bouhelier-Dereux:2012} for $\epsilon_b \ne 1$).
Considering a Drude metal
\begin{equation}
 \epsilon_m=1-\frac{\omega_p^2}{\omega^2+i\Gamma_p \omega} \,.
\end{equation}
the $n^{th}$ resonance occurs for $\epsilon_m(\omega_n)=-\frac{n}{n+1}\epsilon_b$ that is for 
\begin{eqnarray}
\omega_n=\omega_p \sqrt{\frac{n}{2n+1)}}
\end{eqnarray}
so that the $n^{th}$ polarisability becomes
\begin{eqnarray}
\alpha_n&=&\frac{n(\epsilon_S-1)}{n\epsilon_S+(n+1)}R^{2n+1} \,,\\
&=&
\frac{-n\omega_p^2}{(2n+1)(\omega^2+i\Gamma_p \omega)
-n\omega_p^2}R^{2n+1} 
\end{eqnarray}
we now use $n\omega_p^2=(2n+1)\omega_n^2$ to write
\begin{eqnarray}
\alpha_n&=&\frac{-\omega_n^2}{\omega^2-\omega_n^2+i\Gamma_p \omega
}R^{2n+1}\,,\\
&=&\frac{-\omega_n^2}{(\omega-\omega_n)(\omega+\omega_n)+i\Gamma_p \omega}R^{2n+1}
 \end{eqnarray}
Finally, near a resonance, $\omega\approx \omega_n$; we obtain 
\begin{eqnarray}
\alpha_n&\approx &\frac{-\omega_n^2}{2\omega_n(\omega-\omega_n)+i\Gamma_p \omega_n}R^{2n+1} \,,\\
&\approx & \frac{-\omega_n}{2(\omega-\omega_n)+i\Gamma_p}R^{2n+1}
 \end{eqnarray}
Thus $\vert \alpha_n \vert^2$ follows a Lorentzian profile peaked on the $n^{th}$ resonance $\omega_n$ and with a full-width at half maximum (FWHM) 
$\Gamma_p$ associated to Joule losses in the metal.

\subsubsection{Radiative losses.}
\label{sectGrad}
In the previous section, only Joule losses appear although radiative losses are expected, at least for the dipolar (n=1) mode. 
This difficulty comes from the approximation that the electric field (or its $n^{th}$ order gradient for the next modes) is assumed constant over the particle size. 
Taking into account the variation of the electric field over the particle (or, more easily, applying the optical theorem to ensure the energy conservation), it is possible to show that the above expressions are improved using the effective polarisabilities \cite{GCFIJMS:2009,GCF-Derom-Vincent-Bouhelier-Dereux:2012}
\begin{equation}
\alpha_n^{eff}=\left[1 - i~\frac{(n+1)k_b^{2n+1}}{n(2n-1)!!(2n+1)!!}\alpha_n \right ]^{-1}\alpha_n \;,
 \end{equation}
that behave near a resonance as
\begin{eqnarray}
\label{eq:alphaNapprox}
\alpha_n^{eff}&\mathop{\sim}\limits_{\omega_n}&\frac{\omega_n}{2(\omega_n-\omega)-i\Gamma_n} R^{2n+1}\,,\\
\nonumber
\Gamma_n&=&\Gamma_p+\Gamma_n^{rad} \,, \\
\nonumber
\Gamma_n^{rad}&=&\omega_n \frac{(n+1)(k_0R)^{2n+1}}{n(2n-1)!!(2n+1)!!} \,,
\end{eqnarray}
where $\Gamma_n$ is the total decay rate of the $n^{th}$ mode, that includes both ohmic losses and radiative scattering. As expected, for a given mode $n$, the radiative scattering rate $\Gamma_n^{rad}\propto R^{2n+1}$ increases with the particle size since it couples more efficiently to the far-field.  


\subsubsection{Near-field coupling rate.} 
Finally, we assume a dipolar emitter close to the MNP surface. We use the approximate expressions of the Hankel function (Eq. \ref{eq:B_nQS}), the Mie coefficient (Eq. \ref{eq:BnApprox}) and the  quasi-static polarisabilities (Eq. \ref{eq:alphaNapprox}) to express the radial component of the the Green tensor (Eq. \ref{eq:ImGzz}) 

\begin{eqnarray}
\label{eq:Gqs}
G_S^{rr}({\bf r_d},{\bf r_d}) &\approx & \frac{1}{4\pi k_b^2}\sum_{n=1}^\infty  \frac{(n+1)^2R^{2n+1}}{r_d^{2n+4}}\frac{\omega_n}{2(\omega_n-\omega)-i\Gamma_n}
\end{eqnarray}  
so that the near-field coupling rate to LSP$_n$ (Eq. \ref{lien}) is approximated by
\begin{eqnarray}
\vert \kappa_{\omega,n}(\mathbf{r}_d)\vert^2 &=&\frac{k_0^2}{\hbar\pi\epsilon_0}d_{eg}^2 Im\left[G_n^{rr}({\bf r_d},{\bf r_d})\right] \\
&\approx &  \frac{d_{eg}^2}{4\hbar \pi \epsilon_0 \epsilon _b}  \frac{(n+1)^2R^{2n+1}}{r_d^{2n+4}}\frac{\omega_n/2}{(\omega_n-\omega)^2+(\Gamma_n/2)^2}\frac{\Gamma_n}{2}
\end{eqnarray}  
Identification with Eq. (\ref{structuration}) leads to the following expression for the coupling strength
\begin{eqnarray}
g_n \approx  \frac{d_{eg}\sqrt{\omega_n/2\hbar \epsilon_0}}{2 n_b}  \frac{(n+1)R^{n+1/2}}{r_d^{n+2}}
 \end{eqnarray}
 
Finally, from Eq. (\ref{eq:Gqs}), we obtain that the Green tensor follows a first order resonance  
\begin{eqnarray}
G_S^{rr}({\bf r_d},{\bf r_d}) &\approx &\frac{\hbar\epsilon_0}{k_0^2 d_{eg}^2}  \sum_{n=1}^\infty \frac{g_n^2(\mathbf{r}_d)}{(\omega-\omega_n)^2+(\Gamma_n/2)^2} \left[ -(\omega-\omega_n)+i\frac{\Gamma_n}{2} \right] \;.
\label{eq:Gorder1} 
\end{eqnarray}

\section{Far-field spectrum}
\label{sect:FarFieldSPec}
\begin{figure}
\includegraphics[width=8cm]{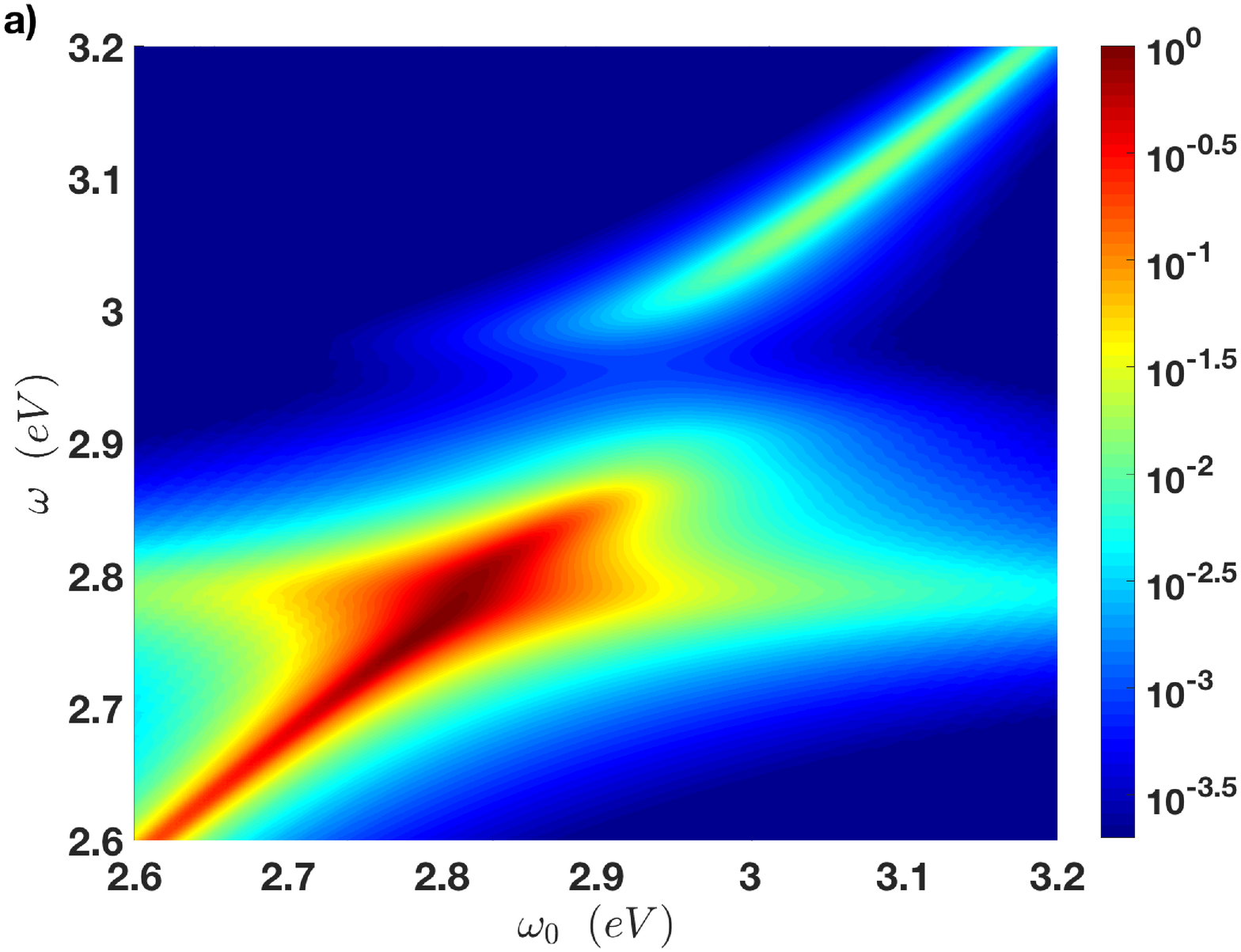}
\includegraphics[width=8cm]{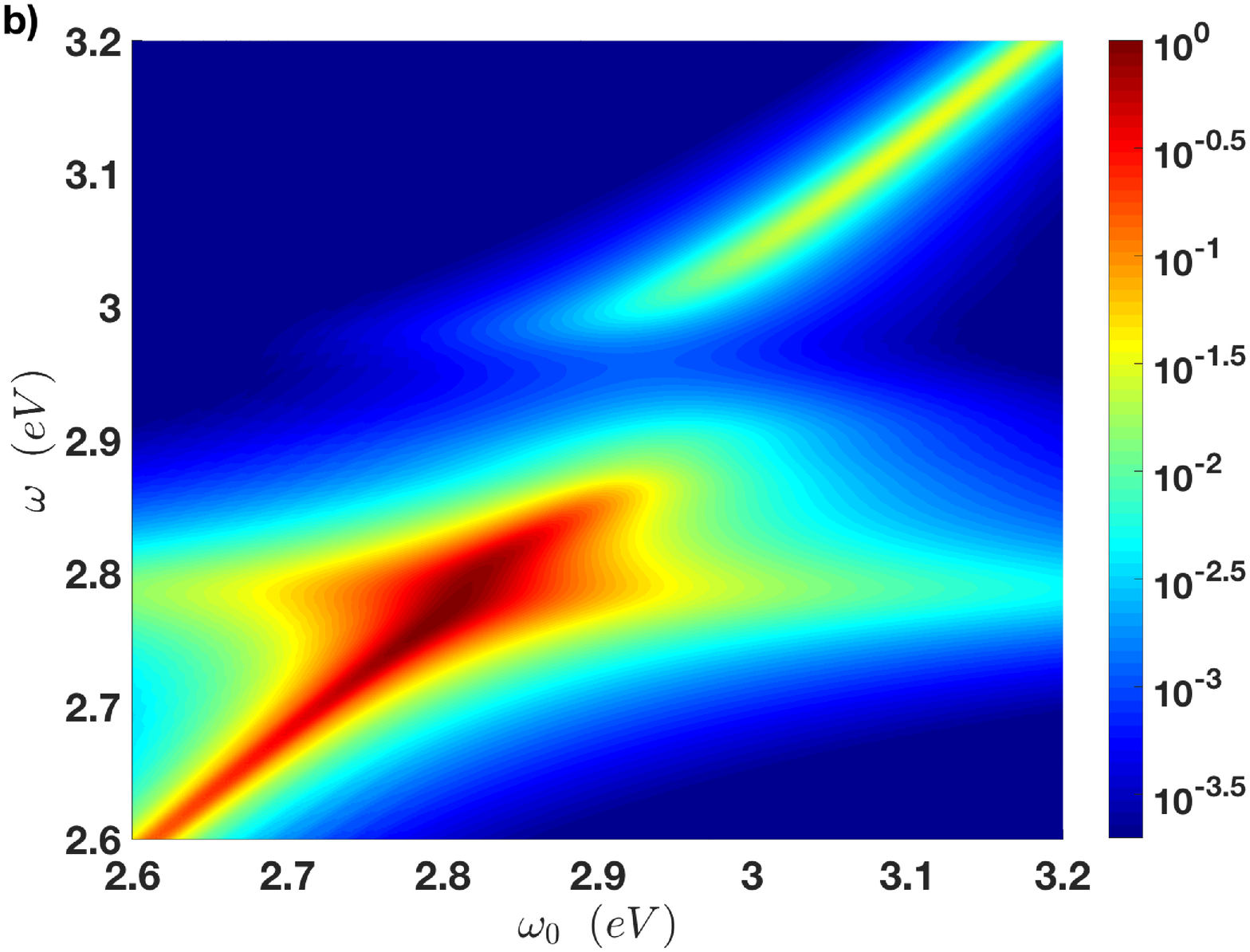}
\caption{Far-field emission. a) Spectrum calculated at the detector position (Eq. \ref{eq:Sfar}) varying the TLS emission angular frequency $\omega_0$. b) LSP$_1$ population $\vert C_1(\omega) \vert^2$.  All figures are normalized with respect to their maximum.} 
\label{fig:FarField2}
\end{figure}

The light spectrum at position $\mathbf{r}_d$ is related to the polarization spectrum by \cite{Hakami-Wang-Zubairy:2014}
\begin{eqnarray}
S(\mathbf{r},\omega)&=&\frac{1}{2\pi}\left|\frac{k_0^2}{\epsilon_0}  \mathbf{G}(\mathbf{r},\mathbf{r}_d,\omega)\cdot \mathbf{d_{eg}}\right|^2 P(\omega) \;.
\label{eq:Sfar}
\end{eqnarray}
$k_0^2  \mathbf{G}(\mathbf{r},\mathbf{r}_d,\omega)\cdot \mathbf{d_{eg}}/\epsilon_0$ describes the electric field scattered at the point $\mathbf{r}$ by a dipolar source located at $\mathbf{r}_d$ so that the far-field spectrum clearly appears as the signal propagating from the hybrid source to the detector position (including both scattering by the MNP and direct free-space propagation). The far-field signal is presented in Fig. \ref{fig:FarField2}a)  when scanning the TLS emission frequency $\omega_0$ \cite{Delga-GarciaVidal:2014}. We observe again a Rabi splitting of $144$ meV for $\omega_0=2.94$ eV that is a reminiscence of the strong coupling regime observed in the polarization spectrum. However, the main contribution comes from the bright LSP$_1$ scattering near $\omega=2.79$ eV (see also Fig. \ref{fig:FarField}). Qualitative understanding of the far-field behaviour can be achieved considering the dipolar LSP$_1$ mode population.
Indeed, we can infer the the total radiated power
\begin{eqnarray}
P_{rad}(\omega) &=&\int S(\mathbf{r},\omega) d\Omega
\end{eqnarray}
and using far-field asymptotic expansion for the Green's tensor, one obtains
\begin{eqnarray}
P_{rad}(\omega) =\frac{1}{2\pi}\gamma^{rad} (\omega) P(\omega)
\label{eq:Prad}
\end{eqnarray}
where $\gamma^{rad}$ is the radiative decay rate (at the angular frequency $\omega$) in presence of the MNP, that depends on the dipole moment orientation \cite{OptExpGCF:2008}. For 
small particle size, the far-field Green's function is peaked at the dipolar LSP$_1$ frequency and the radiative rate is approximated by
\begin{eqnarray}
\gamma^{rad}(\omega) &\approx& n_b\frac{d_{eg}^2\omega^3}{3\pi\epsilon_0 \hbar c^3}\left[1+\frac{4}{d^6}\left \vert \alpha_1(\omega) \right \vert ^2\right]  
\end{eqnarray}
for a radial emitter close to the MNP. 

Since the far-field emission should be governed by the LSP mode radiation, it is of strong interest to express the far-field spectra as a function of the LSP$_n$ population. 
The dynamics of the coupled system is governed by the effective Hamiltonian (\ref{eff_hamil}),  $i\hbar \partial_t \ket{\Psi_{eff}(t)}=\hat H_{eff} \ket{\Psi_{eff}(t)}$. In particular,  
\begin{eqnarray}
\label{eq:dCe}
\dot C_e(t)&=&-\frac{\gamma_0}{2} C_e(t) -\sum_{n=1}^{N}g_nC_n(t) \;, 
\end{eqnarray} 
so that
\begin{eqnarray}
C_e(\omega) &=&\frac{C_e(0)-\sum_{n=1}^N g_n C_n(\omega)}{i(\omega_0-\omega)+\gamma_0/2}  \;,
\\
C_n(\omega)&=&\int_0^\infty dt e^{i(\omega-\omega_0) t}C_n(t)dt
\nonumber
\end{eqnarray} 
and for an emitter initially in its excited state, the far-field radiated signal can be written as
\begin{eqnarray}
P_{rad}&=&\frac{1}{2\pi}\gamma^{rad} P(\omega) \\
\nonumber 
&\approx &n_b\frac{d_{eg}^2\omega^3}{3\pi\epsilon_0 \hbar c^3}\left[1+\frac{4}{d^6}\left \vert \alpha_1(\omega) \right \vert ^2\right]  
\left\vert \frac{1-\sum_{n=1}^N g_n C_n(\omega)}{i(\omega_0-\omega)+\gamma_0/2}   \right\vert^2 \\
&\approx & n_b\frac{d_{eg}^2\omega^3}{3\pi\epsilon_0 \hbar c^3}\left[1+\frac{4}{d^6}\left \vert \alpha_1(\omega) \right \vert ^2\right]  
\frac{\vert1-g_1 C_1(\omega)\vert^2}{(\omega-\omega_0)^2+(\gamma_0/2)^2}
\nonumber 
\end{eqnarray}
since the radiative decay rate selects the dipolar emission near $\omega \approx \omega_1$ for small particles. Finally, one can infer that the radiated power is proportional to the LSP$_1$ population in a rough approximation, $P_{rad}  \sim \vert C_1(\omega)  \vert ^2$. We compare in Fig. \ref{fig:FarField2} (se also Fig. \ref{fig:FarField} in the main text) the far-field emission and  the bright dipolar mode population. We obtain very good agreement, justifying that far-field emission is governed by the dipolar LSP$_1$ mode scattering. We attribute the discrepancy to the fact that $P_{rad}$ describes the scattering in the whole space rather than in a specific direction as $S(\mathbf{r},\omega)$. 

\section{Biorthogonal basis of eigenvectors}
\label{sect:Biorthog}
The aim is to construct the dual basis or left eigenvectors of a non-hermitian Hamiltonian $H$ defined by
\begin{eqnarray}
\langle\Pi_m^L\vert H=\langle\Pi_m^L\vert \lambda_m
\end{eqnarray}
from the right eigenvectors
\begin{eqnarray}
H\vert\Pi_m^{R}\rangle=\lambda_m\vert\Pi_m^{R}\rangle.
\label{eq:RightV}
\end{eqnarray}
We show below that these two sets of vectors satisfy the bi-orthogonality relation
\begin{eqnarray}
\langle\Pi_m^L\vert\Pi_{m'}^{R}\rangle=\delta_{m,m'}.\label{1}
\end{eqnarray}
(\ref{eq:RightV}) defines the non-unitary transformation $T_R$ that diagonalizes $H$
\begin{eqnarray}
T_R^{-1}HT_R=D
\end{eqnarray}
for which the right eigenvectors are on the column
\begin{eqnarray}
T_R=\left[\vert\Pi_1^{R}\rangle \;\;\vert\Pi_2^{R}\rangle \;\;\dots\;\; \vert\Pi_M^{R}\rangle\right].
\end{eqnarray}
The dual basis can be defined from the matrix $T_R$ as
\begin{eqnarray}
T_L^{\dagger}=T_R^{-1}\label{2}
\end{eqnarray}
where the left eigenvectors are on the row of the matrix $T_L^{\dagger}$
\begin{eqnarray}
T_L^{\dagger}=\left(\begin{array}{c}
\langle\Pi_1^L\vert\\
\langle\Pi_2^L\vert \\
\vdots\\
\langle\Pi_M^L\vert \end{array} \right).
\end{eqnarray}

According to the relation \ref{2}, the bi-orthogonality relation \ref{1} is automatically satisfied since 
\begin{eqnarray}
T_L^{\dagger}T_R=T_R^{-1}T_R=\mathds{1}.
\end{eqnarray}
with
\begin{eqnarray}
T_L^{\dagger}HT_R=D.
\end{eqnarray}
We consider now the particular case where the Hamiltonian takes the form
\begin{eqnarray}
H=\left(\begin{array}{ccccc}
0 & \vert g_1\vert \e^{i\theta_1} & \vert g_2\vert \e^{i\theta_2} & \cdots & \vert g_N\vert \e^{i\theta_N}\\
\vert g_1\vert \e^{-i\theta_1} & \Delta_1-i\frac{\gamma_1}{2} & 0 & \cdots & 0\\
\vert g_2\vert \e^{-i\theta_2} & 0 & \Delta_2-i\frac{\gamma_2}{2} & \ddots & \vdots\\
\vdots & \vdots & \ddots & \ddots & 0\\
\vert g_N\vert \e^{-i\theta_N} & 0 & \cdots & 0 & \Delta_N-i\frac{\gamma_N}{2}\end{array} \right).
\end{eqnarray}
One can define a symmetric Hamiltonian from the unitary transformation
\begin{eqnarray}
H_S=S^{\dagger}HS,
\end{eqnarray}
where $S$ is a diagonal matrix of the form
\begin{eqnarray}
S=\left(\begin{array}{ccccc}
1 & 0 & 0 & \cdots & 0\\
0 & \e^{-i\theta_1} & 0 & \cdots & 0\\
0 & 0 & \e^{-i\theta_2} & \ddots & \vdots\\
\vdots & \vdots & \ddots & \ddots & 0\\
0 & 0 & \cdots & 0 & \e^{-i\theta_N}\end{array}\right)\e^{i\kappa}\label{12},
\end{eqnarray} 
with $\kappa$ an arbitrary phase and $S^{\dagger}S=\mathds{1}$. The new Hamiltonian reads
\begin{eqnarray}
H_S=\left(\begin{array}{ccccc}
0 & \vert g_1\vert & \vert g_2\vert & \cdots & \vert g_N\vert\\
\vert g_1\vert & \Delta_1-i\frac{\gamma_1}{2} & 0 & \cdots & 0\\
\vert g_2\vert & 0 & \Delta_2-i\frac{\gamma_2}{2} & \ddots & \vdots\\
\vdots & \vdots & \ddots & \ddots & 0\\
\vert g_N\vert & 0 & \cdots & 0 & \Delta_N-i\frac{\gamma_N}{2}\end{array} \right),
\end{eqnarray}
and the symmetrical property of this Hamiltonian implies that $H_S^t=H_S$ where $t$ denotes the transpose matrix. The left eigenvectors $\vert \psi_m^L\rangle$ of $H_S$ can be expressed in terms of the right eigenvectors
\begin{eqnarray}
\vert \psi_m^L\rangle=\vert \psi_m^R\rangle^*.\label{14}
\end{eqnarray}
Indeed  we can write
\begin{eqnarray}
&H_S^{\dagger}\vert\psi_m^L\rangle=\lambda_m^*\vert\psi_m^L\rangle\\
&H_S^{t}\vert\psi_m^L\rangle^*=\lambda_m\vert\psi_m^L\rangle^*\\
&H_S\vert\psi_m^L\rangle^*=\lambda_m\vert\psi_m^L\rangle^*,
\end{eqnarray}
and make the link with the definition of the right eigenvectors
\begin{eqnarray}
&H_S\vert\psi_m^R\rangle=\lambda_m\vert\psi_m^R\rangle.
\end{eqnarray}
Using the relation (\ref{14}) and the transformation $S$ allowing on to express  the right and left eigenvectors of $H$ in terms of the right and left eigenvectors of $H_S$
\begin{eqnarray}
&\vert\psi_m^R\rangle=S^{\dagger}\vert\Pi_m^R\rangle\\
&\vert\psi_m^L\rangle=S^{\dagger}\vert\Pi_m^L\rangle,
\end{eqnarray}
we obtain the expression of the dual basis in terms of the right eigenvectors of $H$
\begin{eqnarray}
\vert\Pi_m^L\rangle=SS^{t}\vert \Pi_m^R\rangle^*.
\label{eq:LeftVect}
\end{eqnarray}
According to the matrix form of $S$ (see equation \ref{12}), we can write
\begin{eqnarray}
\vert\Pi_m^L\rangle=\e^{2i\kappa}\left(\begin{array}{ccccc}
1 & 0 & 0 & \cdots & 0\\
0 & \e^{-2i\theta_1} & 0 & \cdots & 0\\
0 & 0 & \e^{-2i\theta_2} & \ddots & \vdots\\
\vdots & \vdots & \ddots & \ddots & 0\\
0 & 0 & \cdots & 0 & \e^{-2i\theta_N}\end{array}\right)\vert \Pi_m^R\rangle^*.
\end{eqnarray}
The effective Hamiltonian studied in the article is defined with $\theta_i=\frac{\pi}{2}$ $\forall i$. The matrix form of $SS^{\dagger}$ can be then written as
\begin{eqnarray}
SS^{\dagger}=\left(\begin{array}{ccccc}
-1 & 0 & 0 & \cdots & 0\\
0 & 1 & 0 & \cdots & 0\\
0 & 0 & 1 & \ddots & \vdots\\
\vdots & \vdots & \ddots & \ddots & 0\\
0 & 0 & \cdots & 0 & 1\end{array}\right),
\end{eqnarray}
where we have chosen the phase $\kappa=\frac{\pi}{2}$. This completes the proof of  Eq. (\ref{eq:LeftVect0}).

\section*{References}

\end{document}